\documentclass[fleqn,usenatbib]{mnras}

\usepackage[dvipsnames]{xcolor}

\usepackage{newtxtext,newtxmath}
\usepackage{makecell}
\usepackage[T1]{fontenc}

\DeclareRobustCommand{\VAN}[3]{#2}
\let\VANthebibliography\thebibliography
\def\thebibliography{\DeclareRobustCommand{\VAN}[3]{##3}\VANthebibliography}

\usepackage{graphicx}	% Including figure files
\usepackage{amsmath}	% Advanced maths commands
\usepackage{float}
\usepackage{orcidlink}
\title[Damped Oscillation]{Does DESI prefer Damped Oscillating Dark Energy over Cosmological constant?}

\author[Hussain et al.]{
	Saddam Hussain \orcidlink{0000-0001-6173-6140},$^{1}$\thanks{E-mail:saddamh@zjut.edu.cn}
	Simran Arora \orcidlink{0000-0003-0326-8945},$^{2}$\thanks{E-mail:arora.simran@yukawa.kyoto-u.ac.jp}
	Qiang Wu$^{1}$\thanks{E-mail: wuq@zjut.edu.cn (Corresponding Author)}
	and Tao Zhu \orcidlink{0000-0003-2286-9009}$^{1}$\thanks{E-mail: zhut05@zjut.edu.cn (Corresponding Author)}
	\\
	$^{1}$Institute for Theoretical Physics and Cosmology, Zhejiang University of Technology, Hangzhou 310023, China\\
	$^{2}$Center for Gravitational Physics and Quantum Information, Yukawa Institute for Theoretical Physics, Kyoto University, Kyoto 606-8502, Japan
}

\date{Accepted XXX. Received YYY; in original form ZZZ}

\pubyear{\the\year{}}

\newcommand{\bq}{\begin{equation}}
	\newcommand{\eq}{\end{equation}}
\newcommand{\bqn}{\begin{eqnarray}}
	\newcommand{\eqn}{\end{eqnarray}}

\newcommand{\km}{\ensuremath{\rm km \ s^{-1} Mpc^{-1}}}
\newcommand{\pp}{\partial}

\newcommand{\higl}[1]{{\color{black}#1}}

\renewcommand{\cite}[1]{\citep{#1}}

\begin{document}
	\label{firstpage}
	\pagerange{\pageref{firstpage}--\pageref{lastpage}}
	\maketitle
	
	% Abstract of the paper
	\begin{abstract}
		We investigate a dark-energy equation of state governed by a damped
		harmonic oscillator equation, admitting underdamped, critically
		damped, and overdamped solutions. Confronting the model with Planck
		CMB distance priors, DESI BAO, BBN, cosmic chronometers, and three
		Type~Ia supernova compilations, we find that the data select an
		underdamped solution yielding $H_0 = 70.9 \pm 1.1$ \km\ with
		DES-Dovekie and $H_0 = 72.0^{+1.4}_{-2.1}$ \km\ with Union3,
		 without any local $H_0$ prior. These higher values of $H_0$
			arise along the $\Omega_{\rm m}$--$H_0$ degeneracy direction while
			the sound horizon remains nearly unchanged at
			$r_{\rm d} \simeq 145$~Mpc, indicating that the enhancement of the
			late-time expansion rate is a geometrical effect that does not
			address the early-time calibration of $r_{\rm d}$. In contrast, the
			Pantheon+ compilation selects a near-critically damped solution with
			a prior-limited positive $w_0$ and $H_0 = 66.23 \pm 0.85$ \km,
			highlighting the sensitivity of the model to the low-redshift
			distance information encoded in the different supernova
			compilations. The Bayesian evidence relative to $\Lambda$CDM is inconclusive for the DES-Dovekie and Union3 combinations, whereas Pantheon+ shows a strong preference for the damped-oscillator model, driven by the departure from $w=-1$ at $z\lesssim0.1$.
		
	\end{abstract}

	
	\section{Introduction}
	The discovery of the late-time accelerated expansion of the Universe through
	Type~Ia supernovae observations~\cite{SupernovaCosmologyProject:1998vns,
		SupernovaSearchTeam:1998fmf} has established dark energy as one of the
	central unsolved problems of modern cosmology. The simplest candidate, a
	cosmological constant~$\Lambda$ with equation of state~$w=-1$, faces
	well-known conceptual difficulties including the fine-tuning and coincidence
	problems~\cite{Weinberg:1988cp,Zlatev:1998tr}, motivating dynamical
	alternatives in which~$w_{\rm de}$ evolves with cosmic time. In this Letter
	we introduce a phenomenological dark energy model in which $w_{\rm de}$
	obeys a damped harmonic oscillator equation in $e$-fold time, governed by
	an oscillation frequency~$f$, a damping coefficient~$b$, and an equilibrium
	point~$w_{\rm de}=w_{\rm m}$. {Confronting this model with a combination of datasets} including Cosmic Chronometers (CC), Planck CMB, DESI BAO, BBN, and
	Type~Ia supernovae compilations (Pantheon+, DES-Dovekie, Union3), we find that
	underdamped solutions are observationally preferred and yield
	$H_0 = 70.9\pm 1.1$ \km\ with DES and $H_0 = 72.0^{+1.4}_{-2.1}$ \km\ with Union3, substantially higher than the Pantheon+ result of
	$66.23\pm 0.85$ \km, revealing a striking supernova-dataset dependence of the inferred Hubble constant within oscillatory dark energy models.
	
	Recent analyses by the Dark Energy Spectroscopic Instrument (DESI)
	collaboration have provided new impetus for dynamical dark
	energy~\cite{DESI:2024kob,DESI:2025zgx}. When the dark-energy equation of state is described by the Chevallier--Polarski--Linder (CPL) parametrization $w(a)=w_0+w_a(1-a)$ \cite{Chevallier:2000qy,Linder:2002et}, the combination of DESI BAO with Planck CMB and Type~Ia supernovae yields a preference for dynamical dark energy at the $2.5$--$4.2\sigma$ level, depending on the supernova dataset~\cite{DESI:2025zgx}. However, the inferred dynamics depend sensitively on the adopted parametrization: the constant-$w$ model remains broadly consistent with $\Lambda$CDM, while CPL suggests a possible phantom--nonphantom transition at $z \lesssim 0.5$. This parametrization dependence signals that low-order Taylor expansions of~$w(z)$ may be too restrictive to capture the full complexity of dark-energy dynamics, motivating more general descriptions with richer temporal structure.
	
	{Model-independent and non-parametric reconstructions of $w(z)$ from current observations have reported indications of non-monotonic, potentially oscillatory behaviour at low redshifts~\cite{Zhao:2017cud,Ormondroyd:2025exu,Escamilla:2021uoj,Kessler:2025kju,Hussain:2025nqy}. Early theoretical studies explored oscillatory dark-energy scenarios within scalar-field models through suitably chosen potentials, demonstrating that a dynamical equation of state can emerge while alleviating the cosmic coincidence problem~\cite{Rubano:2003er,Lazkoz:2007mx,Brown:2017osf,Linder:2005dw}. Furthermore, string-theory-inspired frameworks predict ultra-light axion fields with masses of order the present Hubble scale, $H_0$, whose evolution in periodic potentials naturally generates an oscillating equation of state that remains close to $-1$ at early times while developing potentially observable deviations at low redshift~\cite{Arvanitaki:2009fg,Ruchika:2020avj,Caldwell:2005tm}.
		
		Motivated by these developments, a variety of phenomenological oscillatory parametrizations have been proposed to search for signatures of dark-energy oscillations in cosmological data~\cite{Pace:2011kb,Pan:2017zoh,Li:2011dr,Rezaei:2024vtg}. While useful phenomenologically, such parametrizations are typically introduced in an ad hoc manner and do not arise from a common dynamical principle governing the oscillatory evolution. Moreover, describing independently the present-day value of $w(a)$, phantom-divide crossings, and oscillatory dynamics generally requires more than one degree of freedom. Such models are also of interest in the context of the Hubble tension, as departures from $w=-1$ at late times can modify the expansion history and cosmological distance measures, potentially accommodating larger values of $H_0$ inferred from local observations~\cite{Linder:2005dw}.}
	
	We address this gap by proposing that $w_{\rm de}(N)$, where
	$N=\log a = -\log(1+z)$ is the $e$-fold number, satisfies the second-order equation
	\begin{equation}
		\frac{d^2 w_{\rm de}}{dN^2}
		= -f^2\bigl(w_{\rm de}-w_{\rm m}\bigr)
		+ b\,\frac{dw_{\rm de}}{dN},
		\label{eq:osc}
	\end{equation}
	where $f$ sets the oscillation frequency, $b$ controls the damping
	strength, and $w_{\rm m}$ fixes the equilibrium point $w_{\rm de}=w_{\rm m}$. Equation~\eqref{eq:osc} represents a constant-coefficient realization of the general class of second-order evolution equations for $w_{\rm de}$ introduced in Ref.~\cite{Paliathanasis:2025cuc}. We provide the explicit correspondence in Appendix~\ref{app} and emphasize that, owing to the linear structure of the present model, the resulting solution is exact at all times rather than near the equilibrium point.
	This structure has a clear physical interpretation: the damping term drives $w_{\rm de} \rightarrow w_{\rm m}$ at high redshift, where rapid damping suppresses oscillations and the model approaches a cosmological-constant-like behaviour for $w_{\rm m}=-1$, while at low redshift the damping weakens and oscillatory evolution emerges. Depending on whether $b^2 < 4f^2$,
	$b^2 = 4f^2$, or $b^2 > 4f^2$, Eq.~\eqref{eq:osc} admits
	underdamped, critically damped, and overdamped solutions, respectively, 	thereby unifying a broad spectrum of dark-energy behaviours within a 	single dynamical framework. This is qualitatively distinct from
	existing oscillatory ansätze, which impose an oscillatory form but do not encode a damping mechanism that naturally transitions between
	oscillatory and non-oscillatory regimes.
	
	We constrain the model parameters using Bayesian nested sampling
	against six dataset combinations spanning CC, Planck CMB (compressed likelihood), DESI BAO, BBN, growth-rate ($f\sigma_8$, $\sigma_8$), and three independent supernova compilations (Pantheon+, DES-Dovekie, Union3). \higl{Bayesian model comparison via the log-evidence $\ln Z$ shows
		that for $w_{\rm m}=-1.0$ the posteriors concentrate on the
		underdamped regime for all three supernova compilations. For Union3
		and DES-Dovekie the evidence is inconclusive relative to $\Lambda$CDM ($\Delta\ln Z \gtrsim 0.2$), while the model is moderately favored over CPL ($\Delta\ln Z \gtrsim 2.5$). Pantheon+ instead yields strong evidence for the damped-oscillator model over
		both $\Lambda$CDM and CPL ($\Delta\ln Z > 11$ see Tab. \ref{tab:best_fit_val}), but does so with a lower $H_0$ and a positive present-day $w_0$, underscoring the sensitivity of the inferred oscillatory dynamics to
		the choice of supernova compilation.} The matter clustering amplitude remains stable at
	$\sigma_8 \approx 0.768 \pm 0.020$ across all dataset combinations
	and model variants, consistent with large-scale structure observations.
	These results demonstrate that oscillatory dark energy driven by a
	damped harmonic mechanism provides a physically motivated and
	observationally competitive alternative to standard dark energy parametrizations, with supernova compilations emerging as a key source of systematic uncertainty in determining the inferred value of $H_0$.

	\section{Oscillating Equation of State}
	
	We consider the Universe on large scales to be described by a spatially flat, homogeneous, and isotropic Friedmann--Lemaître--Robertson--Walker (FLRW) spacetime, whose line element is given by
	\begin{equation}
		ds^2=-dt^2+a^2(t)d\vec{x}^2 \ ,
	\end{equation}
	where $a(t)$ is the cosmic scale factor and is related to the redshift through $(1+z)=\frac{a_0}{a(t)}$, with $a_0$ denoting its present value. The Universe is assumed to be filled with radiation (r), baryons (b), dark matter (dm), and dark energy (de), all minimally coupled to gravity and independently satisfying the conservation equation
	\begin{equation}
		\frac{d{\rho}_{i}}{dN} = - 3 (1 + w_i)\rho_{i}\ .
	\end{equation}
	Here, $(\rho_i, \  w_i)$ denote the energy density and equation of state, respectively, of the $i^{\rm th}$ cosmic component.
	
	We assume that radiation and baryons obey their standard equations of state, while dark matter is pressureless, $w_{\rm dm}=0$. The dark-energy sector is characterized by a dynamical equation of state $w_{\rm de}(N)$, whose evolution is governed by the oscillatory parametrization given in Eq.~\eqref{eq:osc}. The corresponding dimensionless Hubble parameter, obtained from the Friedmann equation, is
	\begin{equation}
		E \equiv H/H_0 = \sqrt{\Omega_{\rm r} + \Omega_{\rm b} + \Omega_{\rm dm} + \Omega_{\rm de}} \ ,
	\end{equation}
	where $\Omega_{i} \equiv \frac{\kappa^2 \rho_i}{3 H_0^2}$ denotes the time-dependent dimensionless fractional energy density of the $i^{\rm th}$ component. At the present epoch $(N=z=0)$, the above equation reduces to a constraint relation, allowing one of the density parameters to be expressed in terms of the others
	\begin{equation}
		\Omega_{\rm dm_0} = 1 - \left(\Omega_{\rm r_0}+ \Omega_{\rm b_0} + \Omega_{\rm de_0}\right) \ .
	\end{equation}	
	For the oscillatory dark energy equation of state defined in Eq.~\eqref{eq:osc}, two initial conditions are required
	\begin{equation}
		w_{\rm de}(0)=w_0 \ ,
		\qquad
		\left.\frac{dw_{\rm de}}{dN}\right|_{N=0}=w_a\ ,
	\end{equation}
	to determine the evolution of the dark-energy density. The model is therefore characterized by five parameters, $(w_0,w_a,f,b,w_{\rm m})$. The proposed equation of state exhibits underdamped oscillatory behaviour when $\Delta \equiv b^2-4f^2<0$, for which the solution takes the form
	\begin{equation}
		w_{\rm de}(N)=w_{\rm m}+e^{\frac{bN}{2}}
		\left[
		C_1\cos(\omega N)+C_2\sin(\omega N)
		\right]\ , \label{condition_osc}
	\end{equation}
	where $\omega=\frac{\sqrt{4f^2-b^2}}{2}$. The integration constants are determined by the initial conditions. For positive $b$, the oscillatory contribution becomes increasingly suppressed at higher redshifts ($N<0$), with the suppression rate controlled by the magnitude of $b$. Applying the initial conditions yields
	\begin{equation}
		C_1 = w_0 - w_{\rm m}, \quad
		C_2 = -\frac{b w_0 - 2 w_a - b w_{\rm m}}{2 \omega}\ .
	\end{equation}	
	For $\Delta = 0$, corresponding to the critically damped case, the equation of state becomes
	\begin{equation}
		w_{\rm de}(N)=w_{\rm m}+\left(C_1+C_2N\right)e^{\frac{bN}{2}}\ .
	\end{equation}
	This case represents the boundary between oscillatory and non-oscillatory evolution. The integration constants are given by
	\begin{equation}
		C_1 = w_0 - w_{\rm m}, \quad
		C_2 = w_a - \frac{1}{2} b (w_0 - w_{\rm m}) \ .
	\end{equation}	
	For $\Delta>0$, the general solution takes the form
	\begin{equation}
		w_{\rm de}(N)=w_{\rm m}+C_1e^{r_+N}+C_2e^{r_-N}\ ,
	\end{equation}
	where $r_{\pm}=\frac{b\pm\sqrt{b^2-4f^2}}{2}$. In this regime, the damping term dominates over the oscillatory component, resulting in a purely monotonic evolution of the equation of state with no surviving oscillatory behaviour. For $b>0$ and $N\rightarrow -\infty$, the EoS converges to its equilibrium value. The integration constants are
	\begin{equation}
		C_1 = \frac{-w_a+ r_{-} (w_0 - w_{\rm m})}{-\sqrt{b^2-4 f^2}}, \
		C_2 = \frac{w_a- r_{+} (w_0 - w_{\rm m})}{-\sqrt{b^2-4 f^2}}\ .
	\end{equation}	
	For $b>0$, the equation of state approaches the equilibrium value $w_{\rm m}$ in the asymptotic past. Accordingly, throughout this work, we consider three fixed values of $(w_{\rm m})$: (i) $(w_{\rm m}=-1.0)$, corresponding to $w(N\ll0)\to -1$ (cosmological constant), denoted as Model I; (ii) $(w_{\rm m}=-0.8)$, corresponding to $w(N\ll0)\to -0.8$ (quintessence), denoted as Model II; and (iii) $(w_{\rm m}=-1.2)$, corresponding to $w(N\ll0)\to -1.2$ (phantom), denoted as Model III. In addition, we consider a purely oscillatory scenario by setting $(b=0,w_{\rm m}=-1)$, referred to as Model IV.
	
\begin{figure}		
		{	\includegraphics[width=\columnwidth]{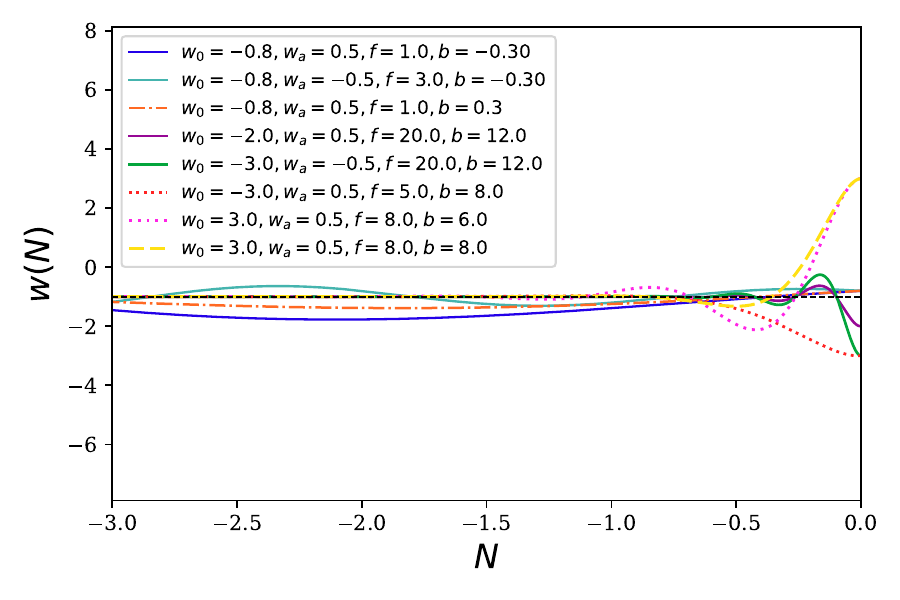} }
		{\includegraphics[width=\columnwidth]{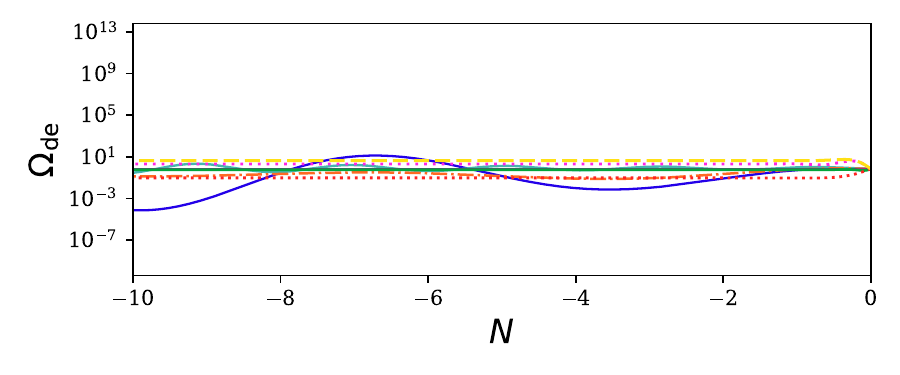}}
		\caption{Numerical evolution of $w(N)$ and the 
			corresponding dark-energy density 
			$\Omega_{\rm de} = \frac{\kappa^2 \rho_{\rm de}(N)}{3H_0^2}$ with respect to e-fold number $N=\ln a$, for different 
			combinations of model parameters with $w_{\rm m} = -1.0$. 
			The figure shows representative choices of the 
			oscillation frequency and damping factor, illustrating 
			underdamped and near-critically damped scenarios in 
			which oscillations emerge at low redshift and are 
			progressively suppressed toward higher redshifts, with 
			the EoS asymptotically approaching $w \to -1$ for 
			$b > 0$.}
		\label{fig:numerical_behav}
	\end{figure}

	The evolution of the equation of state for different parameter choices, 
	together with the corresponding dark-energy density 
	$\Omega_{\rm de}$, is shown in Fig.~\ref{fig:numerical_behav}. 
	The figure illustrates that for positive damping factors, 
	oscillations are efficiently suppressed at high redshifts 
	regardless of the oscillation frequency, and the EoS 
	asymptotically approaches $w \to -1$. Mildly negative 
	values of $b$ lead to slowly growing oscillation amplitudes 
	toward higher redshifts, which may eventually become 
	comparable to the radiation density at early times; this 
	behaviour motivates the lower bound $b > -0.5$ adopted in 
	our analysis. It is worth noting that for $b>0$, the exponential factor grows in the future ($N>0$). Thus, the parametrization is intended to describe the past and present evolution of the Universe. This behaviour is common to phenomenological dark-energy parametrizations, including CPL, and does not affect current observational constraints. 
	The behaviour of the equation of state $w_{\rm de}(z)$ that results from confronting Eq.~\eqref{eq:osc} with the observational datasets (described below) is shown in Fig.~\ref{fig:model1_w_main} for Model I. The oscillations are confined to low redshifts and are rapidly suppressed with increasing $z$, with $w_{\rm de}\to -1$ for $z \gtrsim 0.2$; the model is thus observationally indistinguishable from $\Lambda$CDM at early times. 
	Extrapolated to the future, however, the oscillation amplitude grows
	exponentially and the solution eventually \higl{evolves toward a future cosmological singularity of the big-rip type}, which would
	preclude a physically viable scalar-field realization of the
	parametrization. This future instability is readily cured by promoting the constant damping coefficient to an odd function of $N$,
	$b \to b\tanh(-\gamma N)$, so that Eq.~\eqref{eq:osc} becomes
	\begin{equation}
		\frac{d^{2} w_{\rm de}}{dN^{2}}
		= -f^{2}\bigl(w_{\rm de}-w_{m}\bigr)
		+ b \tanh(-\gamma N)\,\frac{dw_{\rm de}}{dN}\,.
		\label{eq:osc2}
	\end{equation}
	Although this introduces the additional parameter $\gamma$, our parameter
	constraints for Eq.~\eqref{eq:osc} show that the posteriors are insensitive
	to $w_{a}$, which may therefore be fixed without affecting the remaining
	parameters or the overall evolution. One can thus fix $w_{a}$ and constrain
	$\gamma$ instead, leaving the dimensionality of the parameter space
	unchanged. With this modification, $w_{\rm de}(N)\to -1$ as
	$N\to\pm\infty$ for $w_{m}=-1$ and $f\neq 0$, the dark energy
	density approaches a constant, and the late-time evolution asymptotes to a
	stable de Sitter phase. More generally, other choices of $b(N)$ are
	possible, including ones for which the dark energy density may continue to
	grow in the future while remaining oscillatory at the present epoch. Since cosmological observations probe the expansion history only up to the present epoch, we restrict the analysis in this work to
	Eq.~\eqref{eq:osc}; we have verified that Eq.~\eqref{eq:osc2} yields
	closely similar behaviour over the observable range, with mild shifts in
	the inferred parameter values. A detailed exploration of these extensions
	is deferred to future work.

	The jerk parameter $j(z)$ \cite{Rapetti:2006fv} is also shown in Fig.~\ref{fig:model1_w_main}. At low redshift it exhibits damped oscillations about the $\Lambda$CDM value, providing a clear kinematic
	signature of the oscillatory dark-energy component, while at higher redshift all reconstructions rapidly converge to $j = 1$, the value attained by the cosmological constant at all epochs, signaling the restoration of $\Lambda$CDM-like expansion. The amplitude and phase of the low-redshift oscillations depend on the adopted dataset combination, mirroring the behaviour of $w(z)$.

	\section{Observational Samples}
	%	\paragraph*{Observational Samples---}
	To constrain the model parameters, we confront the proposed model with a combination of current cosmological observations. We use 32 Cosmic Chronometer (CC) measurements, including the covariance matrix constructed for the 15 highly correlated samples \cite{Moresco:2012jh,Moresco:2016mzx}, Baryon Acoustic Oscillation (BAO) data from the Dark Energy Spectroscopic Instrument (DESI) Release II \cite{DESI:2025zgx}, and the compressed Planck distance-prior likelihood (Pla), characterized by the set $\{100\Omega_{\rm b}h^2, 100\theta_{*}, R,\Omega_{\rm dm}h^2\}$, where $R$ denotes the shift parameter, $\theta_{*}$ is the angular scale at recombination, and $h \equiv H_0/100$ \cite{Arendse:2019hev}. We further include the Big Bang Nucleosynthesis (BBN) likelihood based on the Primat measurements of primordial abundances\footnote{The data and methodology used to compute the likelihood are available at \url{https://github.com/brinckmann/montepython_public/tree/3.6/data/bbn}.} \cite{Pitrou:2018cgg}.
	
	To investigate the impact of local measurements on the inferred value of the Hubble constant, we also consider the Gaussian prior from the Hubble Space Telescope (HST), commonly referred to as the R19 prior, given by $H_0 = 74.03 \pm 1.42$ km/s/Mpc \cite{Riess:2019cxk}. For Type Ia supernova observations, we employ three independent compilations: Pantheon+ (PP), for which low-redshift data points with $z<0.01$ are excluded \cite{Brout:2022vxf}, the Dark Energy Survey (DES) sample, also known as DES-Dovekie \cite{DES:2025sig}, and the Union3 compilation \cite{Rubin:2023jdq}. The R19 prior is included to assess its effect on the constraints of $H_0$ and its potential role in alleviating the discrepancy with the SH0ES measurement.
	
	In addition, we incorporate redshift-space distortion (RSD) measurements to constrain the amplitude of matter fluctuations, $\sigma_8$. The dataset consists of measurements of $f\sigma_8(z)$ and $\sigma_8(z)$ over a range of redshifts compiled from multiple observational surveys \cite{BOSS:2016wmc}. A complete list of the observational samples used in this analysis can be found in Ref.~\cite{Sahlu:2024pxk}. We define the baseline dataset as CC+DESI BAO+BBN+Pla and perform parameter estimation using the following combinations:
	(i) BASE+PP,
	(ii) BASE+DES,
	(iii) BASE+R19+PP,
	(iv) BASE+R19+DES,
	and (v) BASE+Union3. We further consider the combinations BASE+PP+$\sigma_8+f\sigma_8$ and BASE+DES+$\sigma_8+f\sigma_8$. The joint likelihood is constructed from the total chi-squared through $\mathcal{L}_{\rm tot} \propto \exp(-\chi^2_{\rm tot}/2)$, where
	\begin{equation}
		\chi^2_{\rm tot}= \begin{cases}
			\chi^2_{\rm BASE} + \chi^2_{\rm SN}\ ,\\
			\chi^2_{\rm BASE} + \chi^2_{\rm SN} + \chi^2_{\rm R19}\ ,\\
			\chi^2_{\rm BASE} + \chi^2_{\rm SN} + \chi^{2}_{\sigma_8}+\chi^2_{f\sigma_8}\ .
		\end{cases}
	\end{equation}	
	The posterior distributions of the model parameters are obtained by implementing the model within a Python-based analysis pipeline and performing comprehensive Markov Chain Monte Carlo (MCMC) analyses using the publicly available dynamic nested-sampling package \texttt{dynesty} \cite{Speagle:2019ivv}. The resulting chains are analysed with \texttt{GetDist}\footnote{\href{https://getdist.readthedocs.io/en/latest/}{https://getdist.readthedocs.io/en/latest/}}.
	
	Finally, we compute the Bayesian evidence, quantified by the log-evidence $(\log Z)$, for each model and compare it with that of the baseline flat $\Lambda$CDM cosmology \cite{Speagle:2019ivv}. Following the model-selection criteria adopted in Refs.~\cite{Kass:1995loi}, we interpret the evidence as follows: $0 < |\Delta \log Z| < 1$ indicates inconclusive evidence, $1 < |\Delta \log Z| < 2.5$ corresponds to weak evidence, $2.5 < |\Delta \log Z| < 5$ signifies moderate evidence, and $|\Delta \log Z| > 5$ represents strong evidence.

	\section{Results}

	In the previous section, we outlined the three possible 
	regimes of the oscillatory differential equation. However, 
	to capture the full parameter space in the MCMC analysis, 
	we numerically solve Eq.~\eqref{eq:osc} and determine the 
	corresponding evolution of the dark-energy density 
	$\Omega_{\rm de}$ and the Hubble parameter $H$. Uniform 
	priors are adopted on all model parameters, as listed in 
	Tab.~\ref{tab:prior_range}. These ranges are motivated by 
	preliminary numerical investigations together with physical 
	requirements, such as maintaining a positive dark-energy 
	density, $\Omega_{\rm de}>0$, throughout the cosmic 
	evolution. We further discard solutions that lead to 
	$\Omega_{\rm de}>\Omega_r$ at early times $(N<-8)$ when 
	$b<0$, and consequently impose the lower bound $b>-0.5$. 
	The prior ranges of the remaining parameters are 
	intentionally chosen to be broad, allowing the analysis 
	to explore all physically viable solutions.
	
\begin{figure}
		\resizebox{\columnwidth}{!}{\includegraphics{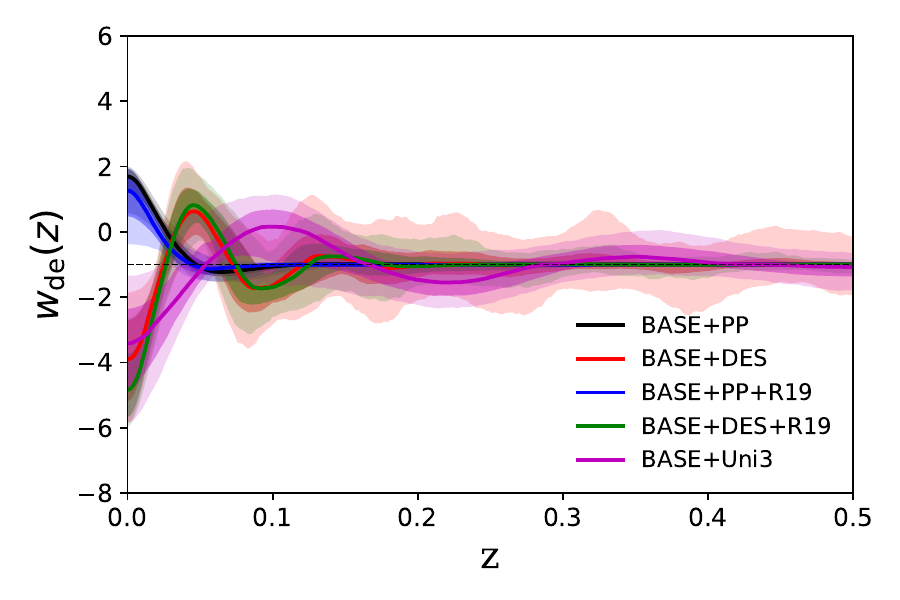} 
		}
		\resizebox{\columnwidth}{!}{\includegraphics{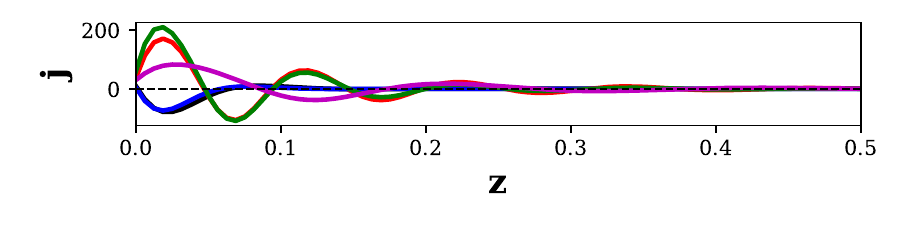} 
		}
		\caption{The evolution of dark energy equation of state $w(z)$ and cosmographic parameter jerk $j(z)$ with respect to redshift $(z)$ for Model~I across 
			distinct dataset combinations. }
		\label{fig:model1_w_main}
	\end{figure}

\begin{figure*}
		\resizebox{\linewidth}{!}{\includegraphics{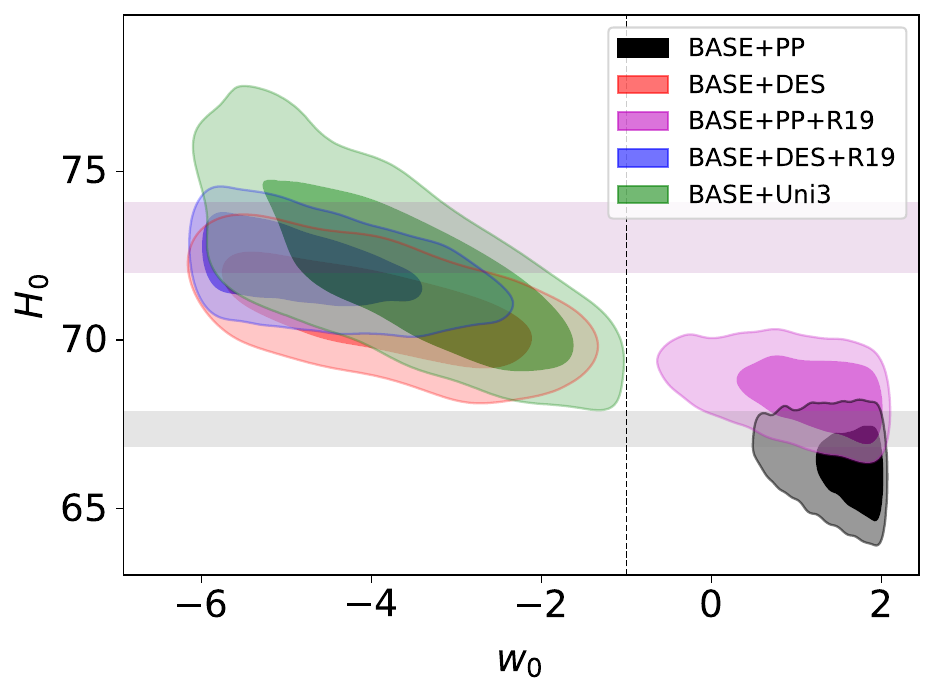}
			\includegraphics{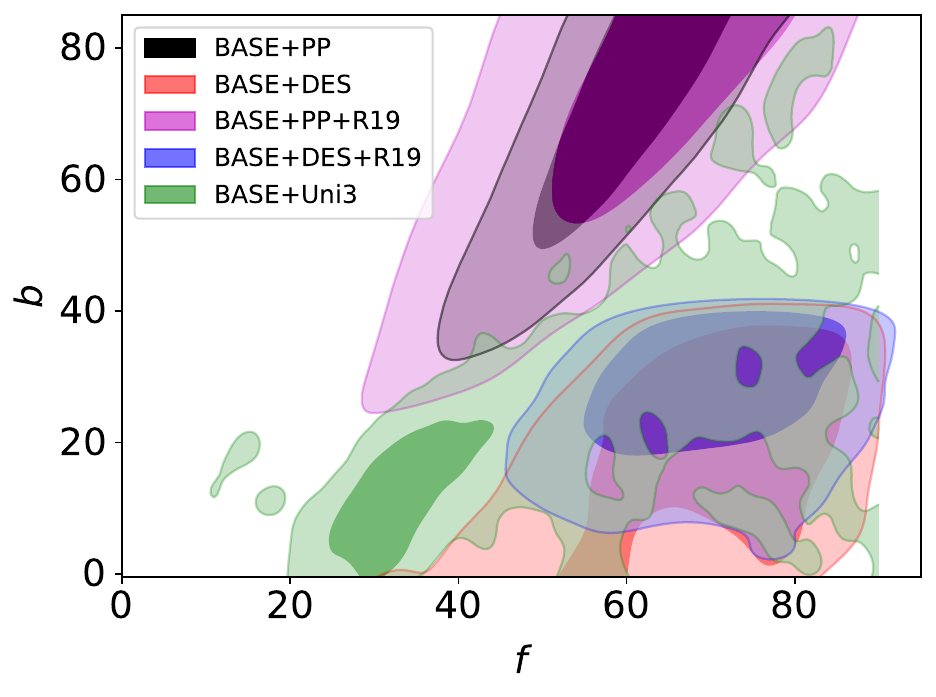}
			\includegraphics{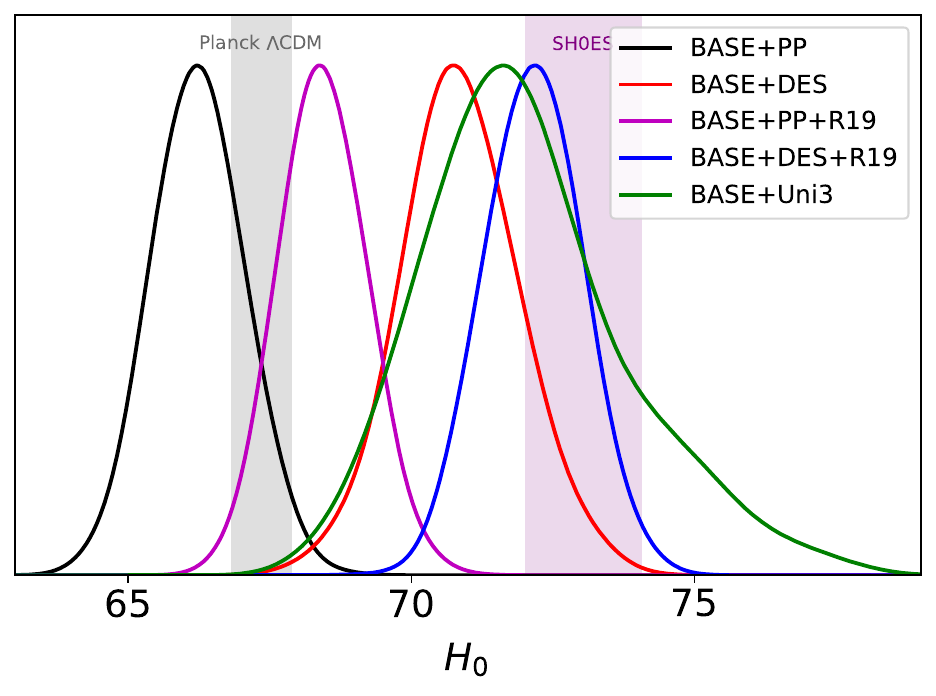}}
		\resizebox{\linewidth}{!}{\includegraphics{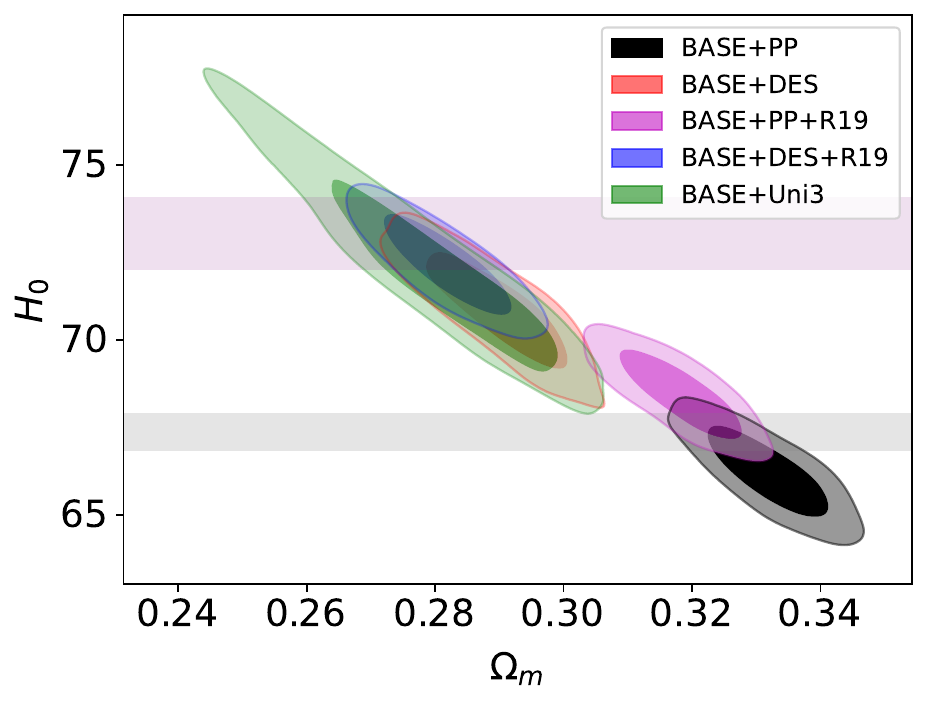}
			\includegraphics{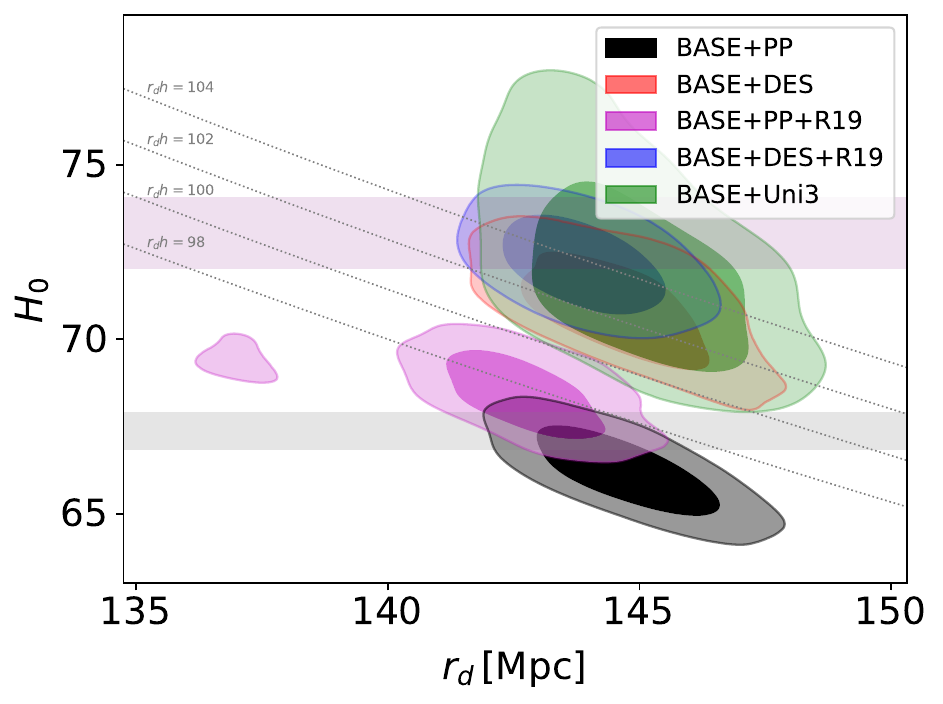}
			\includegraphics{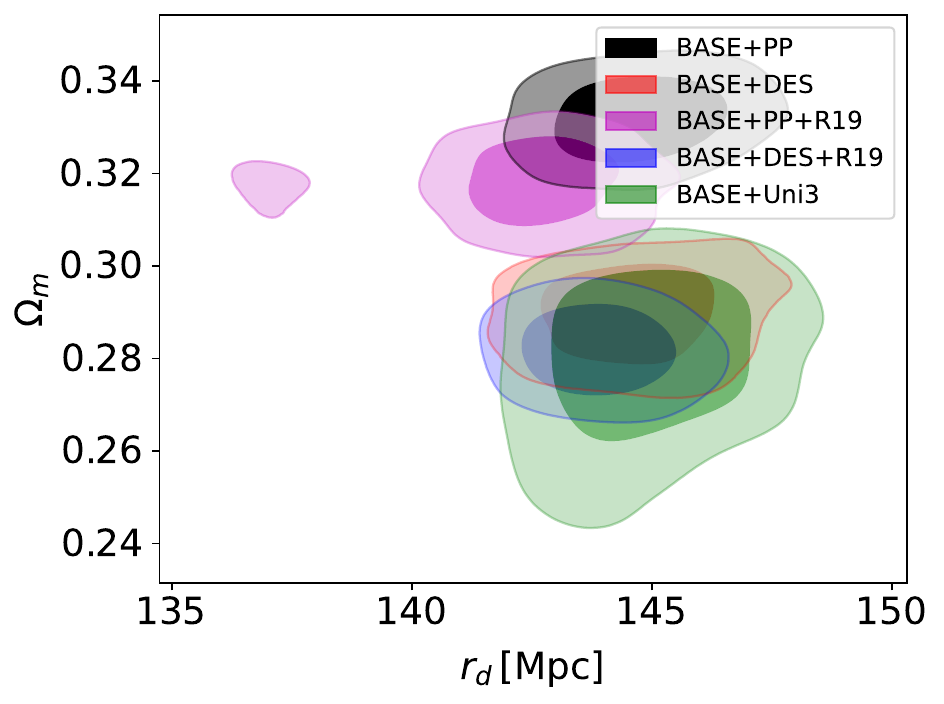}}
		\caption{2-dimensional posterior distributions of $(H_0,w_0)$, $(f,b)$, $(H_0 , \Omega_{\rm m})$, $(H_0, r_d)$ and $(\Omega_{\rm m} , r_d)$ parameters, together with the 1-D posterior distribution of the Hubble constant $(H_0)$, for the different dataset combinations corresponding to Model I. In the left panel, the vertical dashed line marks $w_0=-1$. The Pantheon+ dataset prefers positive values of $w_0$, with the corresponding $H_0$ lying close to the Planck estimate, while for remaining dataset, $w_0<-1$ with $H_0$ close to SH0ES estimated value. The pink and gray shaded bands denote the $68\%$ confidence intervals of the SH0ES and Planck measurements of $H_0$, respectively. Here $\Omega_{\rm m}$ represents the total matter density. A joined posterior plot of $(H_0, \Omega_{\rm m}, r_d, w_0)$ has been displayed in Fig. \ref{fig:joined_h0omrd_plot}.}
		\label{fig:2d_plot}
	\end{figure*}

	The Model~I parameter constraints at the 68\% confidence level are
	reported in Tab.~\ref{tab:best_fit_val} and \ref{tab:best_fit_val2},
	together with those of the CPL parametrization and the baseline
	$\Lambda$CDM model. The corresponding marginalized posteriors are
	shown in Fig.~\ref{fig:model1to4_corner}. For the PP sample we obtain
	$H_0 \simeq 66$--$68$ \km, together with a preference for a large
	positive $w_0>1.2$ (prior-limited; see below) and
	comparatively large values of the frequency and damping parameters,
	placing the system in a near-critically damped regime. As shown in
	Fig.~\ref{fig:model1_w_main}, the resulting equation of state crosses
	the phantom divide $w=-1$ only once, at very low redshift, before
	asymptotically settling at $w=-1$; no oscillatory behaviour survives
	at $z \gtrsim 0.1$ for either the PP or the PP+$H_0$ combination.
	The Bayesian evidence relative to $\Lambda$CDM (displayed in
	Fig.~\ref{fig:H0_best}) is nevertheless substantial,
	$\Delta\ln Z \approx 12$ (PP) and \higl{$\approx 7.7$} (PP+R19),
	indicating a strong preference for a dark-energy component whose
	equation of state departs from $w=-1$ only in the very recent past
	($z \lesssim 0.1$), evolving toward non-accelerating values today.
	This behaviour may arise from the persistent degeneracy between
	$H_0$ and $\Omega_{\rm m}$ as seen in Fig. \ref{fig:2d_plot}, combined with the geometric constraints
	from the supernova data, which favor a transition phase around
	$z \sim 0.05$ and consequently drive the equation of state toward
	positive values. Since we have filtered the Pantheon+ sample by
	imposing $z>0.01$, the ability of the data to constrain the detailed
	evolution of the dark-energy component is reduced. Therefore, it
	remains unclear whether the observed departure reflects
	astrophysical systematics such as progenitor-age evolution, as
	suggested in Refs.~\cite{Sah:2026mgk,Chung:2025cgv}, a
	possibility disputed by the host-galaxy analysis of
	Ref.~\cite{Wiseman:2026nty}, inter-survey calibration
	differences, or a genuine feature of the dark-energy evolution. A
	more comprehensive investigation with additional independent
	supernova samples and improved calibration control is required. We therefore conclude that, for the PP-based combinations, the data favor a late-time departure from $\Lambda$ without exhibiting the sustained oscillatory behaviour characteristic of dynamical dark energy at $z > 0.1$.
	
	In contrast, the DES, DES+R19, and Union3 combinations yield a genuinely
	oscillatory equation of state, as shown in
	Fig.~\ref{fig:model1_w_main}, with the oscillations visible at both the
	68\% and 95\% confidence levels out to $z \sim 0.5$. The oscillation
	amplitude grows toward lower redshift, becoming most pronounced at
	$z \lesssim 0.2$ where the dark energy contribution dominates the
	expansion rate, and drives the present-day equation of state deep into
	the phantom regime, $w_0 \lesssim -3$. This phantom phase, opposite
	in sign to that preferred by the PP sample, raises the inferred
	expansion rate to $H_0 \gtrsim 71~\mathrm{km\,s^{-1}\,Mpc^{-1}}$
	(Fig.~\ref{fig:2d_plot}), thereby reducing the tension with the SH0ES
	measurement, $H_0 = 73.04 \pm 1.04~\mathrm{km\,s^{-1}\,Mpc^{-1}}$, to
	approximately $1.5\sigma$, notably, without the inclusion of any
	local $H_0$ prior. The joint $(w_0, H_0)$ posterior in
	Fig.~\ref{fig:2d_plot} makes the mechanism explicit: $H_0$ increases
	monotonically as $w_0$ becomes more negative along the degeneracy
	direction. The constraint on $w_0 =-3.9\pm 1.1$ tabulated in Tab. \ref{tab:best_fit_val} should, however, be
	interpreted with care: the marginalized posteriors, for all supernova compilations, extend to the edge
	of the adopted prior, $w_0 \in [-6, 2]$, reflecting the fact that
	$w_0 = w_{\rm de}(z=0)$ lies at the boundary of the redshift range
	probed by the data, where the constraining power vanishes. This is
	compounded by the degeneracy between the frequency and damping
	parameters visible in Figs.~\ref{fig:2d_plot} and
	\ref{fig:model1to4_corner}: configurations with larger $f$ and
	correspondingly lower $b$ (with $f > b$) produce nearly identical
	expansion histories over the observed range while differing at
	$z \to 0$. Tighter constraints on $w_0$ could be achieved with
	forthcoming low-redshift surveys, by replacing the CMB distance priors
	with the full Planck likelihood, or within the regularized model of
	Eq.~\eqref{eq:osc2}. Furthermore, the
	increase in $H_0$ is accompanied by a decrease in $\Omega_{\rm m}$
	along the corresponding degeneracy direction, while the sound
	horizon remains nearly unchanged, with
	$r_{\rm d} \simeq 143$--$145$~Mpc (a variation of only
	$\sim 1.5\%$), and $\Omega_{\rm m}h^2$ remains constant at the
	percent level, where $h \equiv H_0/100$. Thus, the model evolves
	primarily along the $\Omega_{\rm m}$--$H_0$ degeneracy direction at
	an approximately fixed sound horizon. The larger values of $H_0$
	inferred from the DES and Union3 combinations therefore originate
	from a late-time geometrical modification rather than a change in
	the early-time physics governing $r_{\rm d}$. Consequently, this
	shift in $H_0$ should not be interpreted as a complete resolution of
	the Hubble tension, but rather as a preference for a higher
	late-time expansion rate in these supernova compilations. This
	interpretation is consistent with recent studies
	\cite{Pedrotti:2025ccw,Huang:2024erq,Cai:2026swf,Sabogal:2026ipu,Adi:2025hyj,Zhang:2025lam},
	which indicate that late-time modifications alone are insufficient
	to fully resolve the $H_0$ discrepancy. Notwithstanding the weak constraint on $w_0$
	itself, the conclusion is robust: in combination with DESI BAO, the
	DES and Union3 supernova samples favor an oscillatory equation of
	state with a phantom excursion at low
	redshift ($z \lesssim 0.1$).

	The Bayesian evidence relative to $\Lambda$CDM remains
	inconclusive for the DES and Union3 datasets in the absence of the
	R19 prior, with $\Delta\ln Z = +0.72$ and $+0.45$, respectively, 
	a verdict consistent with Bayesian reanalyses of the DESI signal
	that find weaker evidence for evolving dark energy than its
	frequentist significance suggests~\cite{Ong:2025utx}. After
	including the R19 prior, the evidence becomes strongly favorable,
	yielding $\Delta\ln Z = +6.12$ for BASE+DES+R19 and $+7.66$ for
	BASE+PP+R19. This indicates that the damped-oscillator model is
	statistically preferred when the local $H_0$ measurement is
	included, although the preference is driven primarily by the tension
	between the supernova-inferred expansion history and the local
	distance-ladder constraint, part of it reflecting the inability of
	$\Lambda$CDM to accommodate the local prior rather than a virtue of
	the model itself. The model constrained with the
	$\rm BASE+DES+f\sigma_8+\sigma_8$ dataset yields
	$\sigma_8 \sim 0.770$, which is consistent with the baseline model.
	Furthermore, the reported Bayesian evidences are not artifacts
	of the broad priors adopted for the model parameters. In
	Appendix~\ref{app:prior_sensitivity}, we perform a prior-sensitivity
	analysis, following the methodology applied to supernova-offset
	models in Ref.~\cite{Lovick:2023tnv}, and demonstrate that the
	inconclusive evidence obtained for the DES and Union3 datasets
	originates primarily from the constraining power of the data rather
	than from any artificial prior-volume effect.
	
	An expanded discussion of Models~II, III, and IV is provided in
	Appendix~\ref{app}. These models are designed as controlled variations
	of Model~I: Models~II and III shift the mean of the oscillation to
	$w_{\rm m} = -0.8$ and $w_{\rm m} = -1.2$, corresponding to asymptotic equations of
	state $w_{\rm de} \to -0.8$ (quintessence-like) and
	$w_{\rm de} \to -1.2$ (phantom-like), respectively, while Model~IV
	isolates the role of damping by setting $b = 0$. All three consistently
	yield $H_0 \simeq 67$--$69~\mathrm{km\,s^{-1}\,Mpc^{-1}}$, and
	Models~II and III exhibit no oscillatory behaviour
	(Fig.~\ref{fig:model1to4_evo_best}).
	
	In Model~II, the present-day equation of state is small and positive
	for the PP and DES combinations, and lies in the range
	$-1 < w_0 < 0$ for DES+R19 and Union3; in all cases $w(N)$ evolves
	toward $-1$ into the past before settling at the asymptotic value
	$-0.8$. Only for Union3 does $w(N)$ cross the phantom divide, remaining
	phantom for $N \lesssim -1.0$ before relaxing to $-0.8$. This evolution
	resembles that of a minimally coupled quintessence field with an
	exponential potential, for which the equation of state stays close to
	$-1$ at intermediate redshifts and grows toward the stiff-fluid value
	$w = 1$ at late times; for Union3 the low-redshift behaviour is instead CPL-like.
	
	In Model~III, $w_0 \gtrsim 0$ for the PP, PP+R19, and DES combinations,
	with $w(N)$ evolving toward $-1$ and crossing the phantom divide near
	$N \sim -0.5$; for DES+R19 and Union3 the equation of state is phantom
	at the present epoch and saturates at the asymptotic value $-1.2$.
	Crucially, this persistent phantom phase does \emph{not} raise $H_0$
	above $70~\mathrm{km\,s^{-1}\,Mpc^{-1}}$ for the same dataset
	combinations that reach $H_0 \gtrsim 71~\mathrm{km\,s^{-1}\,Mpc^{-1}}$
	in Model~I. The comparison thus isolates the origin of the effect: it
	is the damped oscillatory structure of $w(z)$ at low redshift, not a
	phantom equation of state per se, that accommodates the higher
	expansion rate.
	
	Finally, Model~IV removes the dissipative term entirely ($b=0$),
	isolating the purely oscillatory behaviour of the equation of state and
	impose a comparatively narrow flat frequency prior, $0.0 < f \leq 10$. In the absence of damping,
	the preferred frequency remains in the low-frequency regime,
	$f = 4.7 \pm 2.4$ (Tab.~\ref{tab:best_fit_val}), corresponding to a
	slowly oscillating dark-energy fluid whose energy density
	$\rho_{\rm de}$ remains nearly constant at high redshift. For all
	dataset combinations the equation of state lies in the quintessence
	regime at the present epoch, crosses the phantom divide near
	$N \sim -0.5$, and continues to oscillate at higher redshift
	(Fig.~\ref{fig:model1to4_evo_best}). The inferred expansion rate,
	$H_0 \simeq 67$--$69~\mathrm{km\,s^{-1}\,Mpc^{-1}}$, is comparable to that of the CPL parametrization. Model~IV thus demonstrates that
	low-redshift dynamics, including a phantom crossing, is not by
	itself sufficient: an undamped, perpetual oscillation leaves the
	Hubble tension intact. Combined with Models~II and III, this completes
	the picture: neither a smoothly evolving equation of state (II, III)
	nor an undamped oscillation (IV) raises $H_0$; only the under-damped 
	oscillatory structure of Model~I, in which the amplitude grows sharply
	toward $z \to 0$, accommodates
	$H_0 \gtrsim 71~\mathrm{km\,s^{-1}\,Mpc^{-1}}$.

	\section{Conclusion}

	We have investigated a damped harmonic oscillator equation of state
	for dark energy, confronting it with CMB distance priors, DESI BAO,
	BBN, cosmic chronometers, and three supernova compilations. Model~I
	($w_{\rm m} = -1$) yields $H_0 > 70~\mathrm{km\,s^{-1}\,Mpc^{-1}}$ for the DES
	and Union3 compilations, although the Bayesian evidence remains
	inconclusive relative to $\Lambda$CDM. The increase in $H_0$ occurs
	along the $\Omega_{\rm m}$--$H_0$ degeneracy direction while keeping
	$r_d$ nearly fixed at $143$--$145$~Mpc, demonstrating that a purely
	late-time modification alone is insufficient to resolve the $H_0$
	tension. Nevertheless, the model provides a useful demonstration of
	the mechanism required to achieve a higher value of $H_0$, which other
	parameterizations such as CPL fail to reproduce and instead tend to
	remain closer to the Planck-inferred value.
	
	The persistent $\Omega_{\rm m}$--$H_0$ degeneracy remains
	present even after including the R19 prior with the DES compilation.
	In this case, the model attains strong evidence over the baseline
	model ($\Delta\ln Z = +6.12$ for BASE+DES+R19) while yielding
	$H_0 = 72.15 \pm 0.88$~\km, providing an indication that a higher
	late-time expansion rate can naturally emerge within the damped
	oscillatory framework, although we caution that part of this
	preference reflects the inability of $\Lambda$CDM to accommodate the
	local prior. Interestingly, although the evidence relative to
	$\Lambda$CDM remains inconclusive for the same dataset combination
	without the R19 prior, the present model is moderately preferred
	over the CPL parameterization, with $\Delta \ln Z > 2.5$, as shown
	in Fig.~\ref{fig:H0_best}.
	
	The Pantheon+ compilation, in contrast, favors a near-critically
	damped solution with positive $w_0$, yielding
	$H_0 \sim 66.23$~\km\ which indicates a low-redshift geometric
	degeneracy between $H_0$ and $\Omega_{\rm m}$ that is shifted in the
	opposite direction compared with the other supernova compilations.
	The cosmographic jerk parameter $j$ exhibits oscillatory features at
	very low redshift, mirroring those of the dark energy equation of
	state, and saturates to the $\Lambda$CDM value $j = 1$ at redshift
	$z \gtrsim 0.2$, lending further support to the damped oscillatory
	nature. On the contrary, Model~IV, which is oscillatory by
	construction (as $b = 0$), behaves effectively like the baseline
	model and produces $H_0$ close to the Planck value. This model
	does not respond to the low-redshift distance-shape differences
	among the supernova compilations and yields $w_0 \sim -0.9$ for all
	datasets. This further demonstrates the importance of the damping
	factor $b$ introduced in modeling the equation of state, and its
	distinct response to the different supernova compilations.
	
	Overall, the present study provides a framework for investigating oscillatory features of dark energy in a more systematic manner and motivates further analysis using the full CMB likelihood, particularly with the transformed damping factor $b \tanh(-\gamma N)$. This formulation naturally avoids a future big-rip singularity, making it a theoretically viable extension of oscillatory dark-energy models. Moreover, since the model exhibits sensitivity to the geometric information encoded in different supernova compilations, it can provide valuable insights into the ongoing debate regarding possible systematic differences among Pantheon+ and other SN~Ia datasets.

	\section*{Acknowledgments}
	S.H. acknowledges the support of the National Natural Science Foundation of China under Grants No. 12275238, No. 12542053, and No. W2433018, and the National Key Research and Development Program of China under Grant No. 2020YFC2201503. S.A. acknowledges the Japan Society for the Promotion of Science (JSPS) for providing a postdoctoral fellowship during 2024-2026 (JSPS ID No.: P24318). This work of SA is supported by the JSPS KAKENHI grant (Number: 24KF0229).

	\section*{Data availability}
	No new data were generated or analysed in this study. The datasets supporting the findings of this work are publicly available. For a detailed discussion of the data samples and likelihood construction, see Ref.~\cite{Hussain:2025uye}.

	\bibliographystyle{mnras}
	
	\bibliography{ref_prl}

	\appendix

	\section{Parameter Constraints, Marginalized Posteriors, and Cosmological Evolution}
	\label{app}

	Before summarizing our results, we establish the connection
	between Eq.~\eqref{eq:osc} and the general framework of
	Ref.~\cite{Paliathanasis:2025cuc}, in which the equation of state
	satisfies an autonomous second-order law of motion
	$d^2w_{\rm de}/dN^2 \equiv W(w_{\rm de}, w_{\rm de}')$, with the
	cosmological constant $w_{\rm de}=-1$ as an equilibrium point,
	together with a generalization in which the equilibrium is a free
	constant. Throughout this appendix, a prime denotes differentiation
	with respect to $N$. In the present model the equilibrium point is
	$w_{\rm m}$, and the function $W$ takes the constant-coefficient
	form
	\begin{equation}
		W\left(w_{\rm de}, w_{\rm de}'\right)
		= -f^2\bigl(w_{\rm de}-w_{\rm m}\bigr) + b\,w_{\rm de}'\,.
	\end{equation}
	Introducing the shifted variable $\bar{w}=w_{\rm de}-w_{\rm m}$,
	Eq.~\eqref{eq:osc} becomes
	\begin{equation}
		\bar{w}'' = -f^2 \bar{w} + b\,\bar{w}' = W(\bar{w},\bar{w}')\,.
		\label{new_wbar}
	\end{equation}
	In the framework of Ref.~\cite{Paliathanasis:2025cuc}, the dynamics
	near the equilibrium $(\bar{w},\bar{w}')=(0,0)$, where $W(0,0)=0$,
	is governed by the linearization
	\begin{equation}
		W(\bar{w},\bar{w}')
		= \frac{\pp W}{\pp \bar{w}}\bigg|_{\bar{w}'}\,\bar{w}
		+ \frac{\pp W}{\pp \bar{w}'}\bigg|_{\bar{w}}\,\bar{w}'\,,
	\end{equation}
	and comparison with Eq.~\eqref{new_wbar} identifies
	\begin{equation}
		\frac{\pp W}{\pp \bar{w}}\bigg|_{\bar{w}'} = -f^2,\qquad 
		\frac{\pp W}{\pp \bar{w}'} \bigg|_{\bar{w}}=
		\frac{\pp W}{\pp P}\bigg|_{\bar{w}} = b,
	\end{equation}
	and $P \equiv dw_{\rm de}/dN$. Since $W$ is linear with constant coefficient in our model, this identification is exact rather than a near-equilibrium approximation. Therefore, the solution for $\bar{w}$, or equivalently for $w_{\rm de}$, can be written as
	\begin{align}
		&w_{\rm de} = w_{\rm m} + c_1 e^{\Delta_{+} N}
		+ c_2 e^{\Delta_{-} N}\,, \\
		&\text{where}\quad
		\Delta_{\pm} = \frac{1}{2} \left(\frac{\pp W}{\pp P} \bigg|_{\bar{w}} \pm 
		\sqrt{\frac{\pp W}{\pp P} \bigg|_{\bar{w}}^2
			-4\frac{\pp W}{\pp \bar{w}}\bigg|_{\bar{w}'}}\right),
	\end{align}
	holds globally, and the under-, critically-, and over-damped regimes
	of Eq.~\eqref{condition_osc} correspond to the sign of
	$b^{2}-4f^{2}$.

\begin{figure*}
		\resizebox{!}{0.5\textheight}{	
			\begin{tabular}{cc}
				\textbf{Model I} & \textbf{Model I}\\
				\includegraphics[width=.47\textwidth]{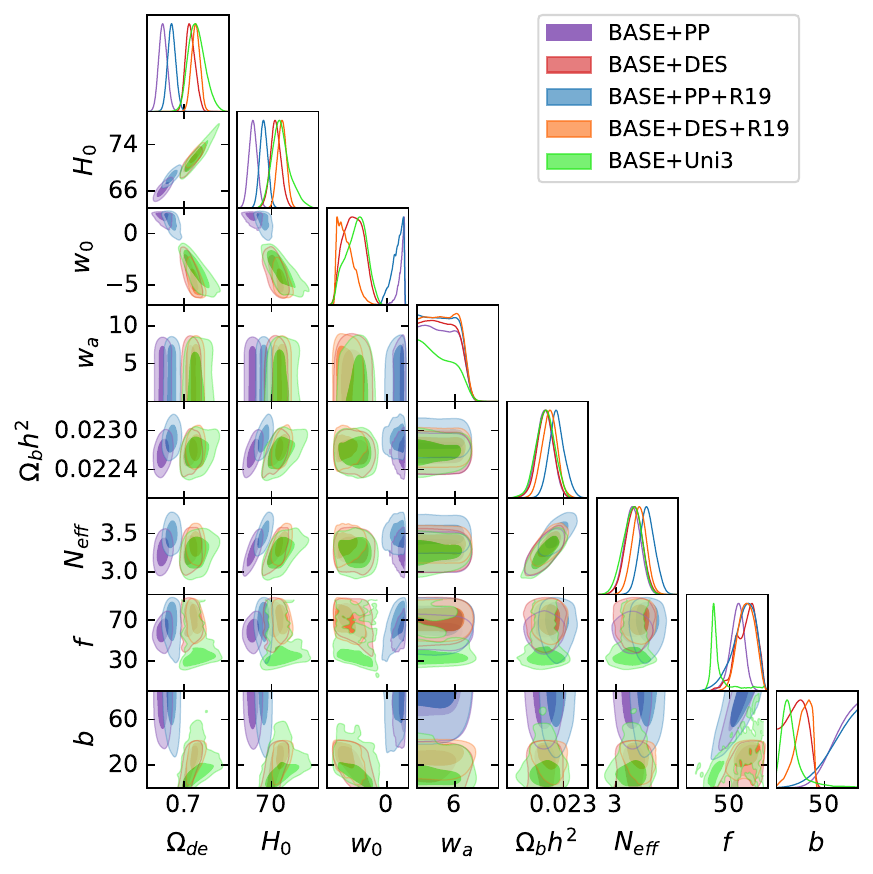} &
				\includegraphics[width=.47\textwidth]{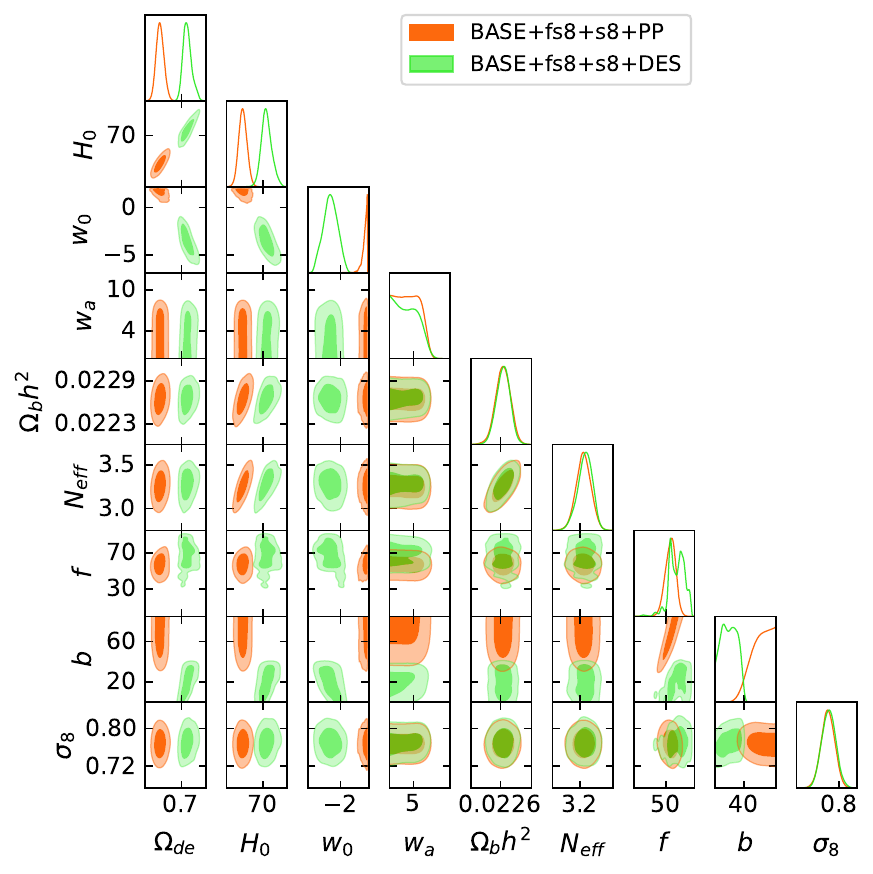}\\
				\textbf{Model II} & \textbf{Model III}\\
				\includegraphics[width=.47\textwidth]{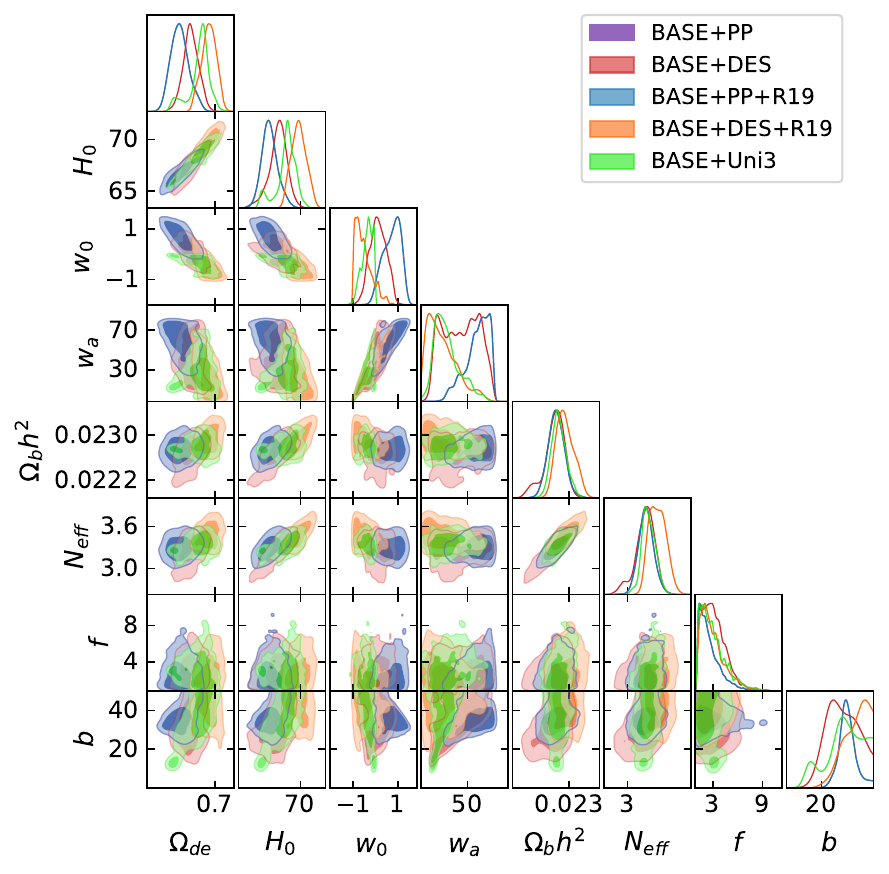} &
				\includegraphics[width=.47\textwidth]{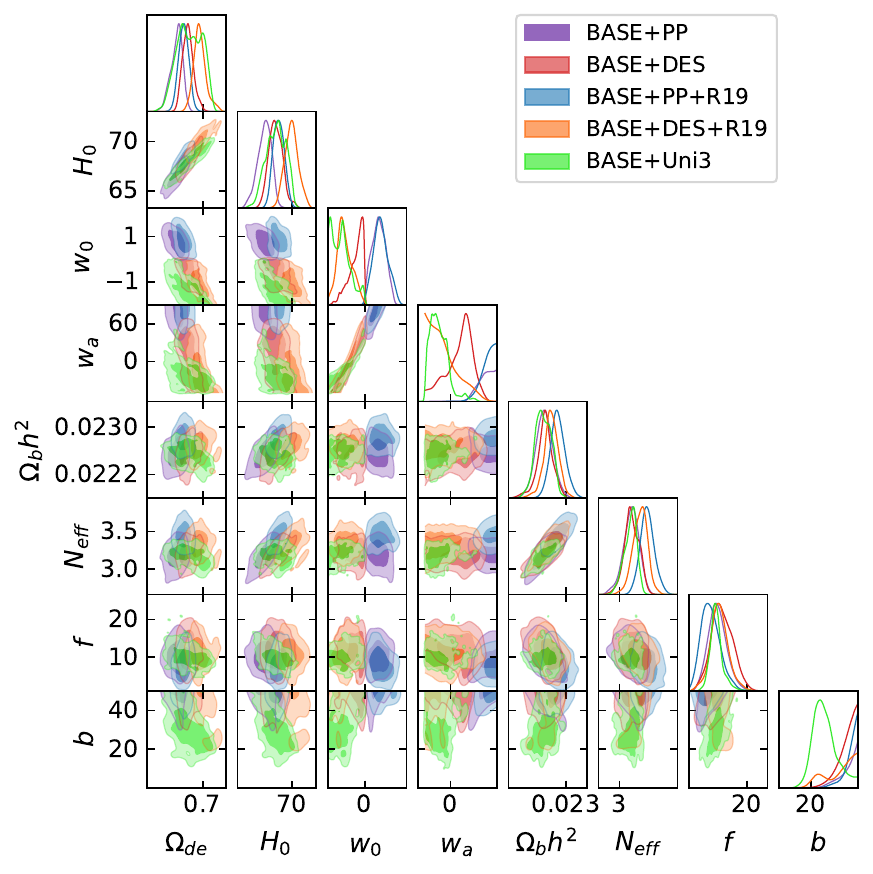}\\
				\textbf{Model IV} & \textbf{Model IV}\\
				\includegraphics[scale=0.47]{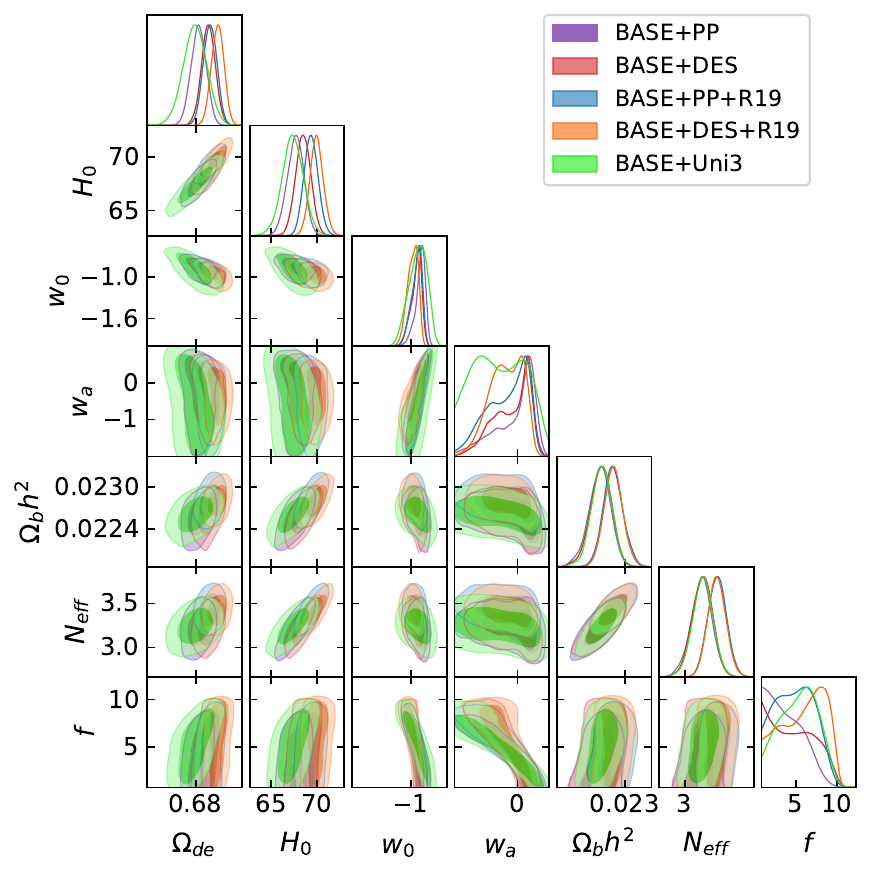} &			
				\includegraphics[scale=0.47]{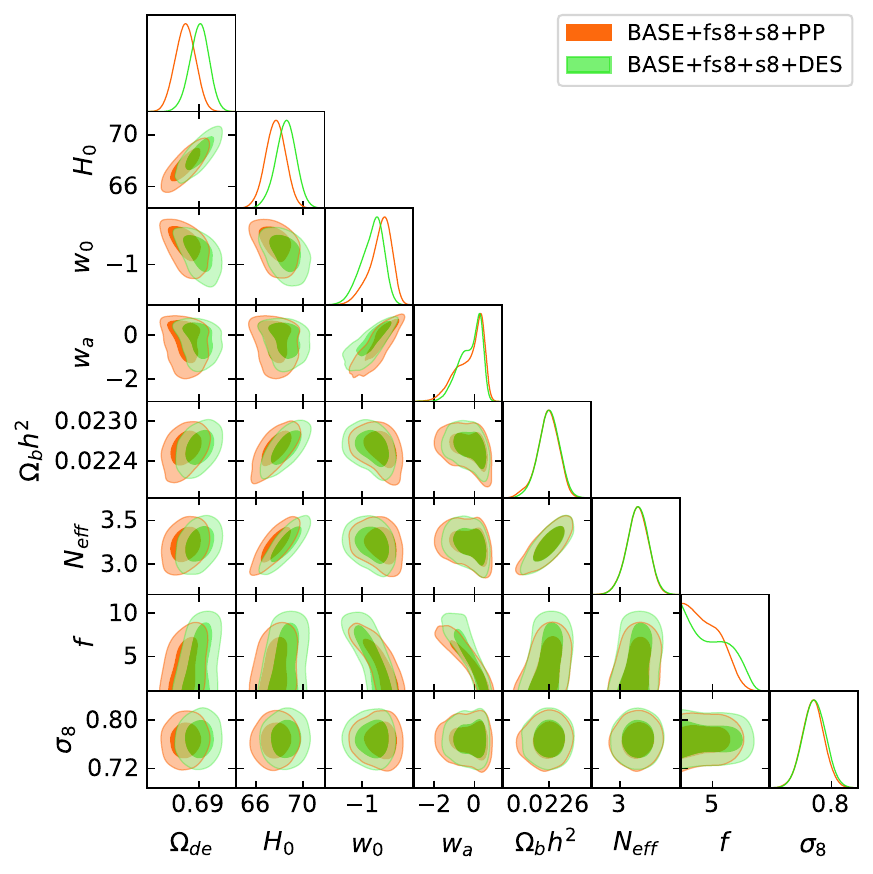}
		\end{tabular}}
		\caption{Marginalized 1D-2D posterior distributions for Model I, II, III  and IV. The corresponding dataset combinations are indicated in the legends of each panel.}
		\label{fig:model1to4_corner}
	\end{figure*}

\begin{figure*}
		\centering
		\resizebox{\linewidth}{!}{
			\includegraphics[scale=0.5]{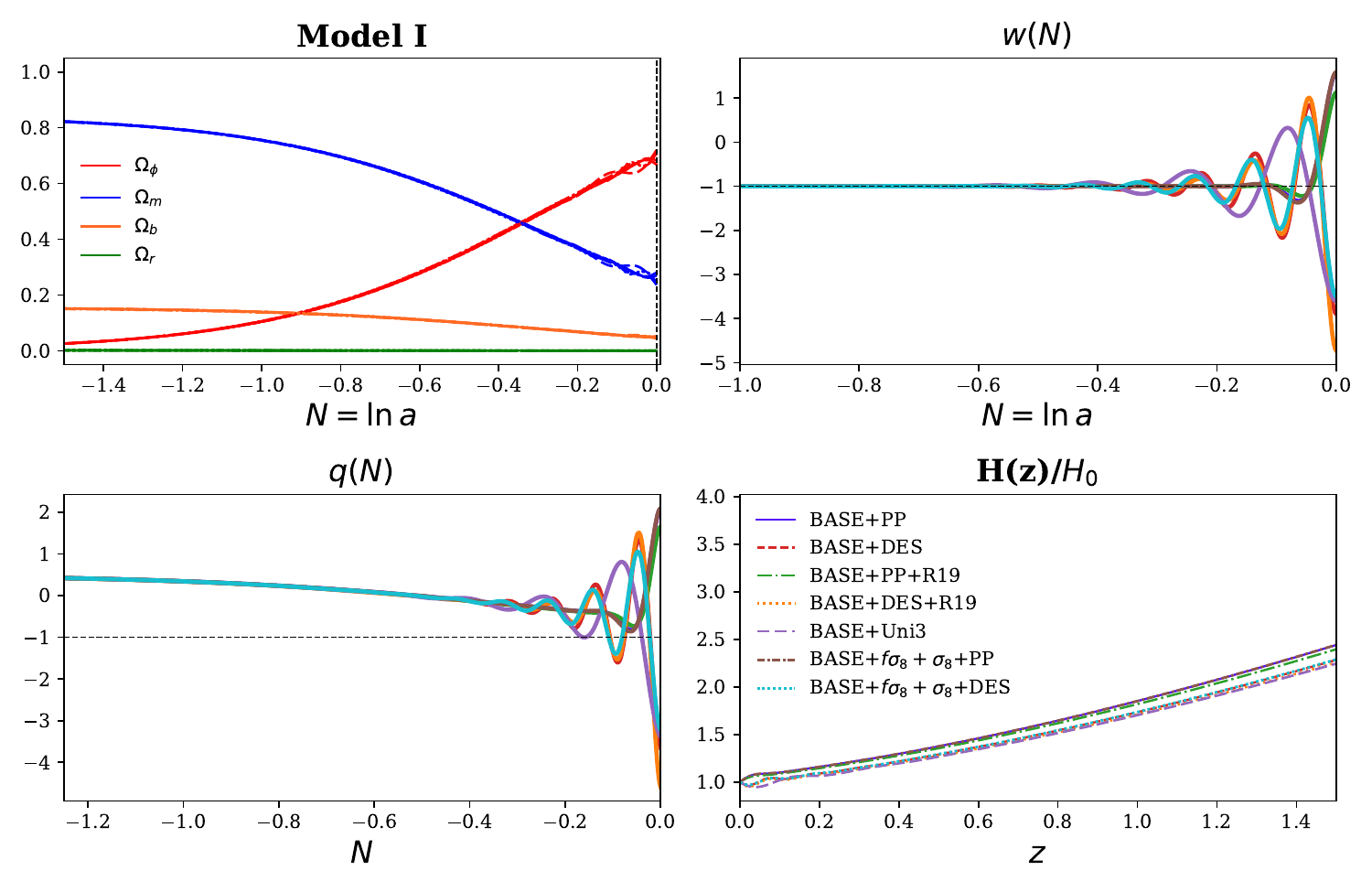}
			\includegraphics[scale=0.5]{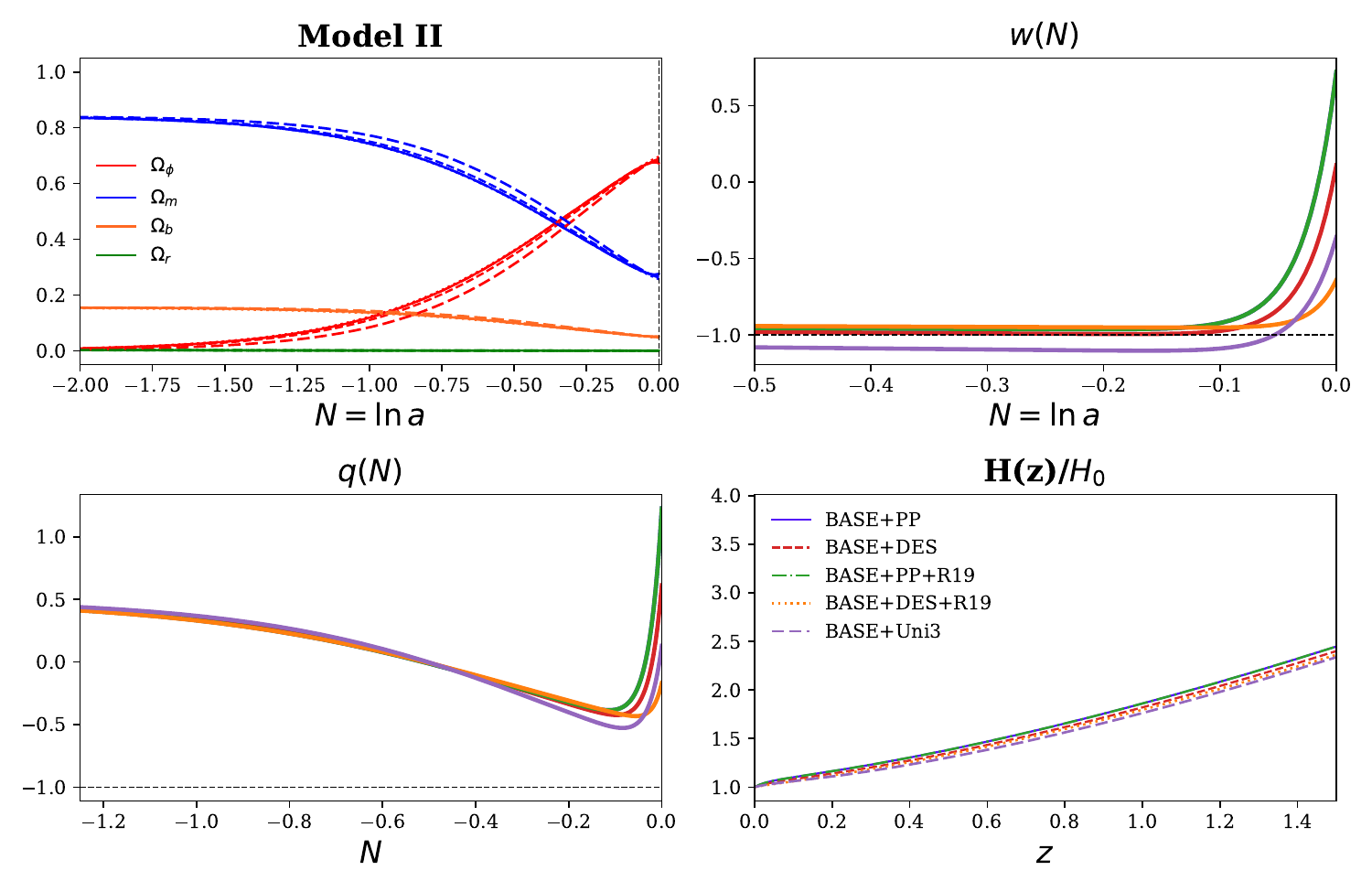}	}
		\resizebox{\linewidth}{!}{
			\includegraphics{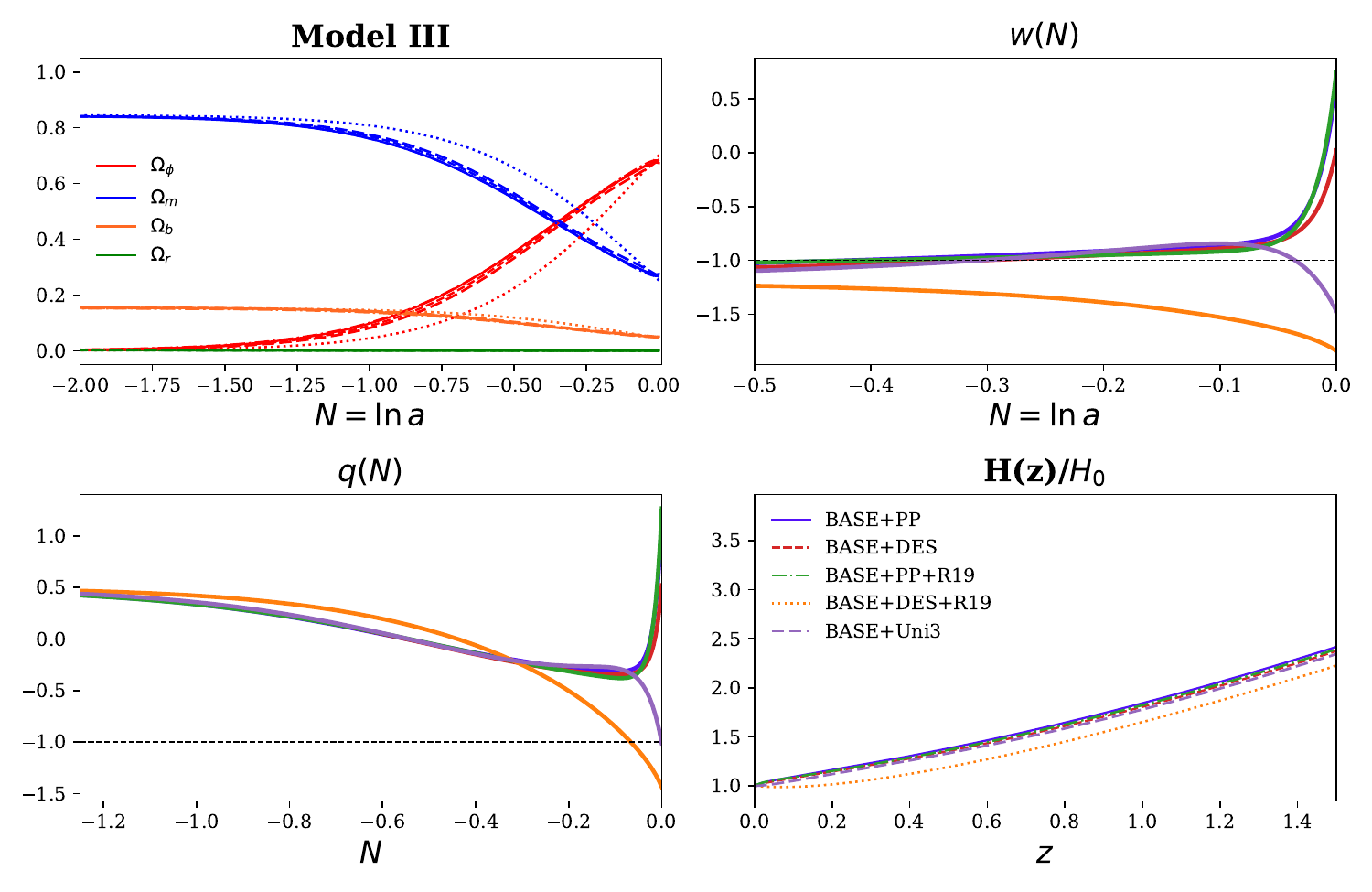}
			\includegraphics{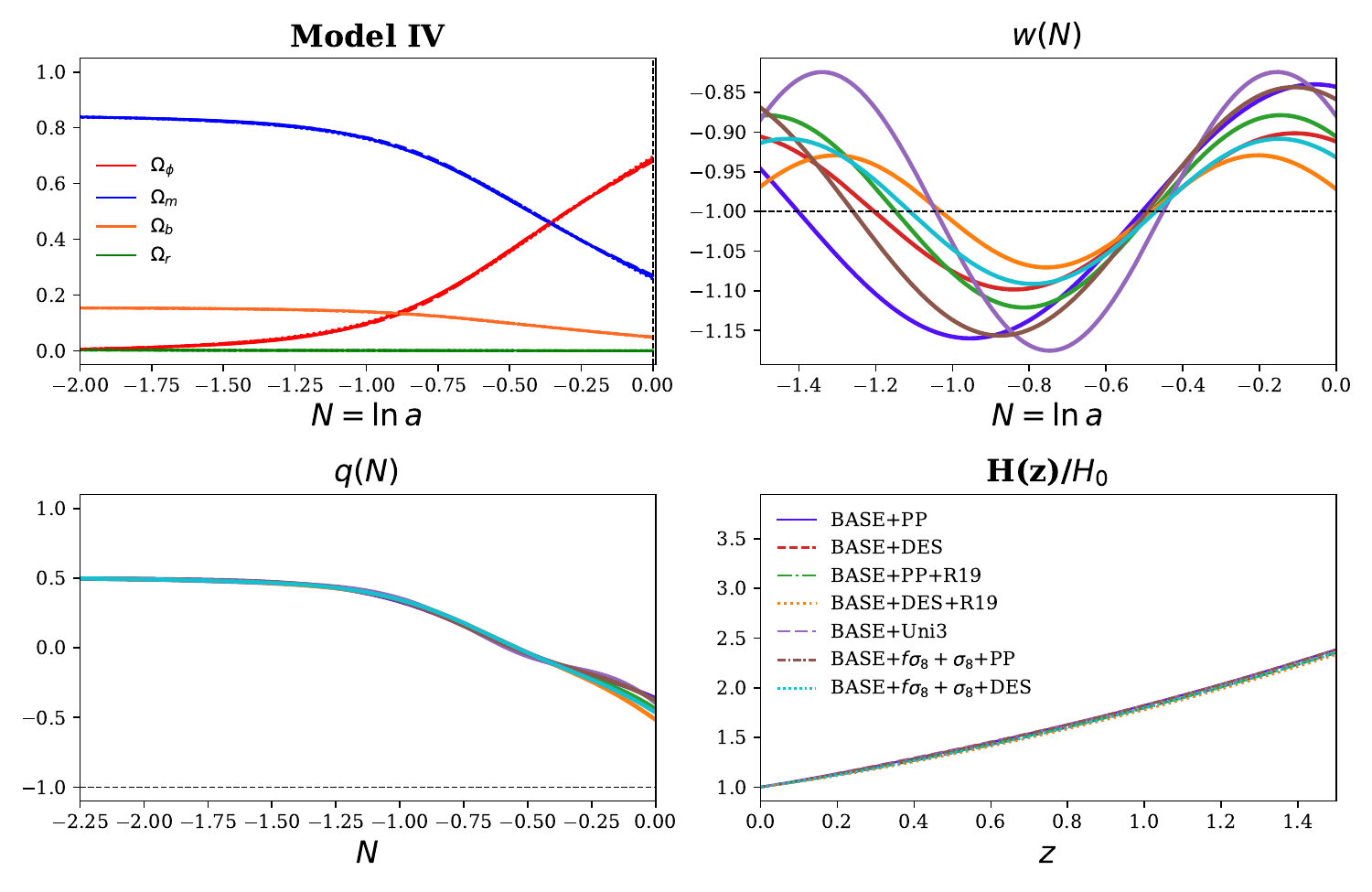}}
		
		\caption{Evolution of the background cosmological parameters for Model~I (\(w_{\rm m}=-1\)) in the top-left, Model~II (\(w_{\rm m}=-0.8\)) in the top-right panel, Model~III (\(w_{\rm m}=-1.2\)) in the bottom-left panel, and Model~IV (\(w_{\rm m}=-1,\; b=0\)) in the bottom-right panel, corresponding to the best-fit values listed in Tables~\ref{tab:best_fit_val} and \ref{tab:best_fit_val2}. Here, $\Omega_{\phi}$ denotes the dark-energy density parameter, $\Omega_{\rm de}$. The dimensionless density parameters are normalized with respect to $H(t)$ for all species according to $\Omega_i=\frac{\kappa^2\rho_i}{3H^2}$. In addition, for each model we plot the dark-energy equation of state $w(N)$, the deceleration parameter $q(N)=-1-\frac{\dot{H}}{H^2}$, and the normalized Hubble parameter $H/H_0$ as functions of either the $e$-fold number $N$ or the redshift $z$.
		}
		\label{fig:model1to4_evo_best}
	\end{figure*}

	Here we summarize the results of the parameter inference.
	Fig.~\ref{fig:model1to4_corner} presents the marginalized one- and
	two-dimensional posterior distributions for Models~I--IV, obtained
	from the various combinations of early- and late-time cosmological
	observations. The corresponding 68\% confidence intervals are listed
	in Tabs.~\ref{tab:best_fit_val}--\ref{tab:best_fit_val2}, and the
	reconstructed evolution of the cosmological quantities at the
	estimated parameter values is shown in
	Fig.~\ref{fig:model1to4_evo_best}. Distinct supernova samples are
	combined with the remaining datasets (as indicated in each table) in
	order to tighten the constraints and break parameter degeneracies. 
	
	For Model~I, the posterior of $w_a$ is unconstrained: fixing it to any
	value within the prior range leaves the remaining parameters
	unaffected. For the other models, $w_a$ acquires a well-defined
	(Gaussian or moderately non-Gaussian) posterior and is positively
	correlated with $w_0$. In all models $H_0$ is negatively correlated
	with $w_0$; the correlation is by far the strongest for Model~I
	(Fig.~\ref{fig:2d_plot}), consistent with its role in raising $H_0$,
	whereas it remains mild for Models~II--IV
	(Fig.~\ref{fig:model1to4_corner}). We additionally plot the
	deceleration parameter $q(N)$ in Fig.~\ref{fig:model1to4_evo_best},
	which characterizes the kinematics of the system: it exhibits
	oscillations at low redshift for the DES and Union3 compilations,
	whereas no oscillatory behaviour is recorded for Pantheon+, \higl{clearly reflecting the sensitivity of the inferred constraints to the choice of supernova compilation.}. At higher
	redshift, $q$ saturates to $0.5$ for all dataset combinations,
	recovering the matter-dominated era. Furthermore, the evolution of
	the density parameters, $\Omega_{i} \equiv \rho_{i}/(3H^{2})$, shows
	no significant deviation for any of the datasets, apart from minute
	departures at very low redshift.
	
	{
		Models~II and III display mutually consistent behaviour in the
		frequency and damping parameters: both prefer $b \gg f$, placing the
		system deep in the overdamped regime. As a result, the oscillations
		are strongly suppressed and the equation of state relaxes
		monotonically toward its mean value, as seen in
		Fig.~\ref{fig:model1to4_evo_best} --- in contrast to Model~I
		($w_{\rm m} = -1$), for which an underdamped solution is selected. Even upon
		inclusion of the R19 prior, neither model yields an appreciable
		increase in $H_0$, which remains consistent with the Planck
		$\Lambda$CDM value. The deceleration parameter registers a deviation
		at low redshift for both models before saturating to $0.5$; in both
		cases, $q$ traces the evolution of the dark energy equation of state
		$w(N)$ in a very near redshift regime.
	}
	
	{
		For Model~IV, in which the damping factor vanishes ($b=0$), the
		equation of state is oscillatory by construction rather than by
		preference of the data. Owing to the simplicity of this construction,
		all model parameters are well constrained within their prior ranges:
		$w_0$ is localized around $-1$ and $w_a > -1$, in contrast to the
		prior-limited $w_0$ of Model~I. For all dataset combinations the
		model yields $H_0 \simeq 68$~\km, even after the inclusion of the R19
		prior, and its Bayesian evidence is lower than that of Model~I.
		Moreover, although the preferred oscillation frequency is low, no
		oscillatory signature survives in the deceleration parameter $q(z)$
		(Fig.~\ref{fig:model1to4_evo_best}). The evolution of the density
		parameters remains the same across all considered datasets; only a
		minor shift in $q(0)$, arising from the different values of $w_0$,
		can be visually identified. The Hubble evolution likewise remains
		identical across the choice of datasets. Taken together, these
		results demonstrate that an undamped oscillation neither raises $H_0$
		nor imprints itself on the expansion kinematics $q$.
	}
	
	{
		To assess whether these models remain compatible with the observed
		growth of cosmic structure, we extend the analysis of Models~I and IV
		to include measurements of the growth-rate observable
		$f\sigma_8(z)$. We do not repeat this exercise for Models~II and III,
		which exhibit neither oscillatory behaviour nor an improved Bayesian
		evidence over the $\Lambda$CDM baseline for the background datasets.
		The resulting constraints are tabulated in
		Tab.~\ref{tab:best_fit_val2}, alongside those of the baseline and CPL
		models; we obtain $\sigma_8 \simeq 0.767$ for both models. All
		remaining parameters are consistent with the background-only results
		of Tab.~\ref{tab:best_fit_val}, indicating that the growth data
		neither degrade the quality of the fit nor pull the background
		parameters --- the oscillatory solutions preferred by the supernova
		combinations are fully compatible with the measured growth history.
	}
	
	\begin{figure}
		\resizebox{\columnwidth}{!}{\includegraphics{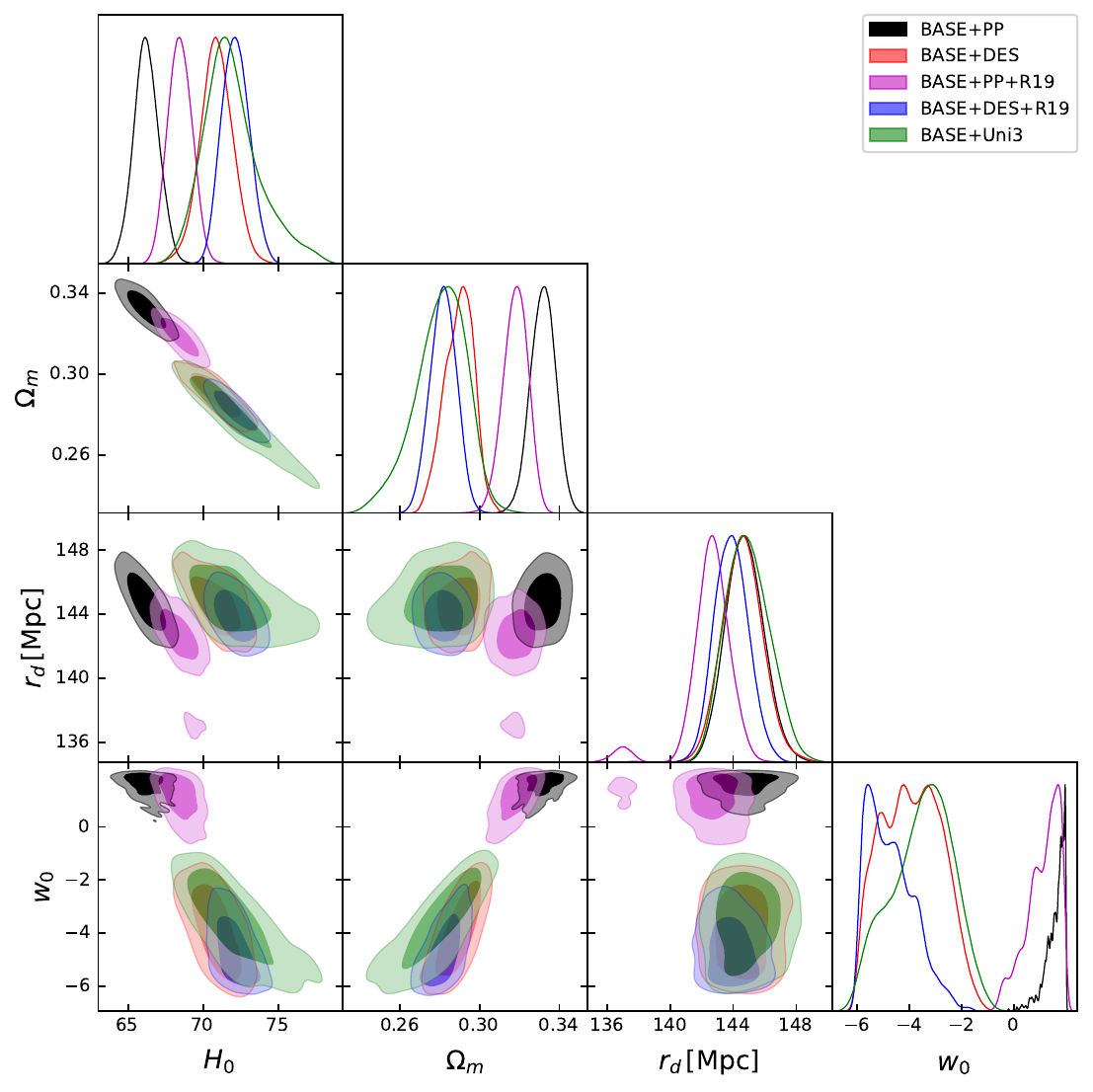}}
		\caption{The marginalized posterior distribution of the parameters $(H_0, \Omega_{\rm m}, r_d, w_0)$ to display the degeneracy among the parameters for Model I with the datasets labeled in the top corner.}
		\label{fig:joined_h0omrd_plot}
	\end{figure}
	
	%=====================================================================
	\subsection{Prior sensitivity of the Bayesian evidence}
	\label{app:prior_sensitivity}
	
	The Bayes factors reported in Tabs.~\ref{tab:best_fit_val} and
	\ref{tab:best_fit_val2} are computed with deliberately wide uniform priors on
	the model parameters ($w_0$, $w_a$, $f$, $b$; see
	Tab.~\ref{tab:prior_range}). Since the Bayesian evidence averages the
	likelihood over the prior volume, a prior substantially wider than the
	posterior support can suppress the evidence through the associated Occam
	factor. Consequently, an ``inconclusive'' verdict may, in principle, reflect
	prior-volume effects rather than the constraining power of the data
	\cite{Ormondroyd:2025phk}. Here we quantify this effect directly using the
	converged posterior samples.
	
	For a uniform prior on a box $\mathcal{B}$ of volume $V$, the evidence is
	$\mathcal{Z}=V^{-1}\int_{\mathcal{B}}\mathcal{L}(\theta)d\theta$. If the
	analysis were instead performed with a narrower box
	$\mathcal{B}'\subset\mathcal{B}$ of volume $V'$, the corresponding evidence
	would be
	\begin{equation}
		\ln\mathcal{Z}'=\ln\mathcal{Z}
		+\ln P(\theta\in\mathcal{B}')
		+\ln\frac{V}{V'} ,
		\label{eq:priorshift}
	\end{equation}
	where $P(\theta\in\mathcal{B}')$ denotes the posterior mass contained within
	the narrower region, evaluated using the original posterior obtained with the
	wide prior. The two correction terms represent the competing effects of
	posterior truncation and prior-volume compression, respectively. When no prior
	reduction is applied, $\mathcal{B}'=\mathcal{B}$, both terms vanish and the
	original evidence is recovered.
	
	The correction is applied directly to $\Delta\ln\mathcal{Z}$ relative to
	$\Lambda$CDM, whose evidence does not involve these additional parameters. A
	fully unconstrained parameter direction does not lead to an evidence gain,
	because the posterior mass scales proportionally with the retained prior
	volume,
	$P(\theta\in\mathcal{B}')\simeq V'/V$, causing the two terms in
	Eq.~\eqref{eq:priorshift} to cancel. We verify this behaviour explicitly for
	$w_a$: halving its prior range changes $\ln\mathcal{Z}$ by less than $+0.1$ for
	all considered datasets in Model I, confirming that $w_a$ remains
	unconstrained and does not influence the model comparison.
	
	We evaluate the evidence shift by imposing data-informed restrictions on the
	priors of $(w_0,f,b)$, chosen from the $[0.1\%,99.9\%]$ posterior quantiles and
	padded by $5\%$ of the original prior width. This procedure provides a
	conservative estimate of the maximum evidence improvement supported by the
	posterior samples, rather than a modification of the priors adopted in the
	main analysis.

	\begin{table*}
		\centering	
		\begin{minipage}{0.58\textwidth}
			\centering
			\begin{tabular}{lcccc}
				\hline
				Parameters & {\boldmath $w_{\rm m} =-1.0$} & {\boldmath $w_{\rm m} = -0.8$} & {\boldmath $w_{\rm m} = -1.2$} & {\boldmath $(w_{\rm m}=-1, b=0)$} \\
				\hline
				\hline
				$\Omega_{\rm de}$ & $[0.2,1.0] $&& &\\
				$H_0$ & $[30,100]$ && &\\
				$N_{\rm eff}$ & $[2.0,4.0]$ && & \\
				$\Omega_{\rm b}h^2$ & $[0,1.0]$&&&  \\
				$w_0$ & $[-6.0, 2.0]$ & $[-4.0,6.0]$& $[-8.0, 10.0]$& $[-2,3]$ \\
				$w_a$ & $[-8.0, 8.0]$ & $[-20, 80]$& $[-50.0, 130.0]$ & $[-4,6]$\\
				$f$ &$[0.0, 90.0]$ & $[0.1, 20.0]$& $[0.1, 90.0]$ & $[0.1, 10]$\\
				$b$ & $[-0.48, 120]$ & $[-0.48,75 ]$& $[-0.48, 95]$ & 0\\
				\hline
				\hline
			\end{tabular}
			\caption{Uniform priors have been adopted for the all the model parameters. Here $h \equiv H_0/100$. }
			\label{tab:prior_range}
		\end{minipage}
	\end{table*}

	For DES and Union3, the resulting change in evidence is
	$\Delta\ln\mathcal{Z}\simeq+0.1$, demonstrating that the inconclusive Bayes
	factors are not driven by excessive prior volume. This behaviour can be
	understood from the posterior distributions of $(w_0,f,b)$ shown in
	Fig.~\ref{fig:2d_plot}, where the parameters extend across most of the original
	prior range along the degeneracy direction, leaving little redundant prior
	volume. For Pantheon+, the shift is larger,
	$\Delta\ln\mathcal{Z}\sim+1.6$, indicating that the original broad prior
	introduced a moderate Occam penalty, with approximately $99\%$ of the posterior
	mass located at $w_0>0$. When the R19 prior is included with PP and DES, the
	shift is $\Delta\ln\mathcal{Z}\sim+1.0$, showing that the original Bayes
	factors are conservative estimates.

	\begin{table*}
		\centering
		\resizebox{\textwidth}{!}{
			\begin{tabular}{l ccccc cc c r}
				\hline
				\multicolumn{10}{c}{\bf Data set: CC+PLA+DBAO+BBN+PP}\\
				\hline
				Model & \(H_0\) & $\Omega_{\rm de}$ & \(N_{\rm eff}\) & $\Omega_{\rm b}h^2$ & \(w_0\) & \(w_a\) & \(f\) & $b$ & $\log Z$\\
				\hline
				\hline
				{\bf I, \((w_{\rm m} =-1.0)\)} & $66.23\pm 0.85$ & $0.6680\pm 0.0063$ & $3.26\pm 0.13$ & \makecell{$0.02265$\\ $
					\pm 0.00015$} & $1.58^{+0.42}_{-0.14}$ & $< 2.32$ & $59^{+9}_{-7}$ & $> 64.3 $ & $-873.04$\\
				
				{\bf II, \((w_{\rm m} = -0.8)\)} & $66.68^{+0.73}_{-0.94}$ & $0.6726^{+0.0061}_{-0.0079}$ & $3.30\pm 0.12$ & \makecell{$0.02274$\\ $
					\pm 0.00016$} & $0.72^{+0.48}_{-0.31}$ & $61.0 ^{+20}_{-6}$ & $2.66^{+0.32}_{-1.6}$ & $36.3^{+3.1}_{-6.5}$& $-880.783$ \\
				
				{\bf III, \((w_{\rm m} = -1.2)\)} & $66.60^{+1.0}_{-0.74}$ & $0.6731^{+0.0074}_{-0.0051}$ & $3.19^{+0.14}_{-0.12}$ & \makecell{$0.02255$\\ $
					\pm 0.00018$} & $0.65^{+0.53}_{-0.36}$ & $91.0 ^{+40}_{-30}$ & $10.20^{+2.7}_{-3.4}$ & $64.0 \pm 10$ & $-881.50$ \\
				
				{\bf IV, \((w_{\rm m} = -1.0, b=0)\)} & $67.65^{+0.94}_{-0.85}$ & $0.6819 \pm 0.0060$ & $3.22^{+0.15}_{-0.13}$ & \makecell{$0.02257$\\ $
					^{+0.00019}_{-0.00017}$} & $-0.843^{+0.10}_{-0.054}$ & $-0.11 ^{+0.81}_{-0.52}$ & $< 4.61$ & --& $ -883.62$ \\

				{\bf \(w_0 w_a\)CDM} & $67.93 \pm 0.90$& $0.6801 \pm 0.0060$& $3.18 \pm 0.15$& \makecell{$0.02252 $ \\ $^{+0.00020}_{-0.00017}$}& $-0.751 \pm 0.055$& $-0.76^{0.24}_{-0.21}$& & &  $-884.37$ \\
				
				{\bf $\Lambda$CDM} & $68.49 \pm 0.29$ & $0.6966 \pm 0.0036$ & & $0.02245\pm 0.00011$ &&&&& $-884.80$ \\
				
				\hline
				
				\multicolumn{10}{c}{\bf Data set: CC+PLA+DBAO+BBN+DES}\\
				\hline
				\hline
				{\bf I, \((w_{\rm m} =-1.0)\)} &  $70.9\pm 1.1 $ & $0.7101^{+0.0063}_{-0.0082}$  & $3.28 \pm 0.13$  & \makecell{$0.02270$\\ $^{+0.00017}_{-0.00015}$} & $-3.9\pm 1.1$ & $< 2.76  $ & $70^{+10}_{-10} $ & $20^{+14}_{-11}$ & $-859.02$\\
				
				{\bf II, \((w_{\rm m} = -0.8)\)} &  $67.6^{+1.1}_{-0.75}$ & $0.6813\pm {0.0062}$ & $3.28^{+ 0.12}_{-0.11}$ & \makecell{$0.02270$\\ $
					^{+0.00023}_{0.00013}$} & $0.11\pm 0.36$ & $42.0 \pm 30.0$ & $3.10^{+0.68}_{-1.9}$ & $38.0^{+8}_{-20}$& $-866.035$\\
				
				{\bf III, \((w_{\rm m} = -1.2)\)} & $67.63 \pm 0.96$ & $0.6814\pm {0.0070}$ & $3.20 \pm 0.13$ & \makecell{$0.02257$\\ $
					\pm 0.00017$} & $0.02^{+0.68}_{-0.55}$ & $52.0 ^{+50}_{-40}$ & $11.3^{+3.0}_{-4.1}$ & $63^{+10}_{-20}$& $-865.86$\\
				
				{\bf IV, \((w_{\rm m} = -1.0, b=0)\)} & $68.49\pm 0.83$ & $0.6897 \pm 0.0055$ & $3.23 \pm 0.14$ & \makecell{$0.02259$\\ $
					^{+0.00019}_{-0.00016}$} & $-0.912^{+0.11}_{-0.061}$ & $-0.19 ^{+0.73}_{-0.48}$ & $< 6.02  $ & --& $ -863.18$ \\

				{\bf \(w_0 w_a\)CDM} & $68.17 \pm 0.89$& $0.6872 \pm 0.0056$& $3.19 \pm 0.14$& \makecell{$0.02253 $ \\ $^{+0.00019}_{-0.00017}$}& $-0.821 \pm 0.057$& $-0.58^{0.25}_{-0.21}$& & &  $-864.54$\\
				
				{\bf $\Lambda$CDM} & $68.55 \pm 0.28$ & $0.6974 \pm 0.0036$ & & $0.02246\pm 0.00011$ &&&&& $-859.74$ \\
				\hline
				\multicolumn{10}{c}{\bf Data set: CC+PLA+DBAO+BBN+PP+R19}\\
				\hline
				\hline
				{\bf I, \((w_{\rm m} =-1.0)\)} & $68.44\pm 0.80 $ & $0.6815\pm 0.0061$ & $3.49 \pm 0.12$ &  \makecell{$0.02288$\\$^{+0.00013}_{-0.00015}$} & $1.13^{+0.84}_{-0.36}  $ & $< 2.70$ & $67^{+20}_{-9}   $ & $> 75.6 $ & $-884.51$ \\
				
				{\bf II, \((w_{\rm m} = -0.8)\)} & $66.68^{+0.73}_{-0.94}$ & $0.6726^{+0.0061}_{-0.0079}$ & $3.30\pm 0.12$ & \makecell{$0.02274$\\ $
					\pm 0.00016$} & $0.72^{+0.48}_{-0.31}$ & $61.0 ^{+20}_{-6}$ & $2.66^{+0.32}_{-1.6}$ & $36.3^{+3.1}_{-6.5}$& $-891.77 $\\
				
				{\bf III, \((w_{\rm m} = -1.2)\)} & $68.40 \pm 0.69$ & $0.6807^{+0.0054}_{-0.0062}$ & $3.44 \pm 0.12$ & \makecell{$0.02282$\\ $
					^{+0.00014}_{-0.00016}$} & $0.75^{+0.55}_{-0.40}$ & $97.0 ^{+40}_{-20}$ & $8.0^{+2.7}_{-3.7}$ & $60\pm 10$ & $-890.02$\\
				
				{\bf IV, \((w_{\rm m} = -1.0, b=0)\)} & $69.35\pm 0.76$ & $0.6910 \pm 0.0054$ & $3.243 \pm 0.12$ & \makecell{$0.02280$\\ $
					^{+0.00015}_{-0.00017}$} & $-0.906^{+0.11}_{-0.066}$ & $-0.36 ^{+0.85}_{-0.51}$ & $4.7 \pm 2.4$ & --& $-890.70 $ \\

				{\bf \(w_0 w_a\)CDM} & $69.23 \pm 0.75$& $0.6898 \pm 0.0053$& $3.40 \pm 0.13$& \makecell{$0.02274 \pm 0.00017$ }& $-0.808 \pm 0.053$& $-0.60^{0.23}_{-0.20}$& & & $-892.70$ \\
				
				{\bf $\Lambda$CDM} & $68.71 ^{+0.29}_{0.26}$ & $0.6993 \pm 0.0035$ & & $0.02250\pm 0.00011$ &&&&& $-892.17$ \\
				
				\hline
				\multicolumn{10}{c}{\bf Data set: CC+PLA+DBAO+BBN+DES+R19}\\
				\hline
				\hline
				{\bf I, \((w_{\rm m} =-1.0)\)} & $72.15\pm 0.88 $ & $0.7181\pm 0.0063$ &$3.36 \pm 0.11$ & \makecell{$0.02274$\\$\pm 0.00014$} & $-4.73^{+0.45}_{-1.2}$ & $0.2\pm 4.6  $ & $70^{+10}_{-9}$ & $27^{+10}_{-4}$ & $-860.68$ \\
				
				{\bf II, \((w_{\rm m} = -0.8)\)} &  $69.92 \pm 0.83$ & $0.6960^{+0.0065}_{-0.0055}$ & $3.48^{+0.12}_{-0.15}$ & \makecell{$0.02292$\\ $
					^{+0.00014}_{-0.00021}$} & $-0.64^{+0.25}_{-0.50}$ & $16.0 ^{+14}_{-24}$ & $3.12^{+0.57}_{-2.0}$ & $51.0^{+20}_{-10}$& $-871.78 $\\
				
				{\bf III, \((w_{\rm m} = -1.2)\)} & $69.94\pm 0.86 $ & $0.7021^{+0.0020}_{-0.014}$ & $3.28^{+0.20}_{-0.095}$ & \makecell{$0.02261$\\ $
					^{+ 0.00028}_{-0.000097}$} & $-1.14\pm 0.49$ & $< -5.12   $ & $10.9^{+2.6}_{-3.0}$ & $70^{+20}_{-9}$& $-873.51$\\
				
				{\bf IV, \((w_{\rm m} = -1.0, b=0)\)} & $69.97\pm 0.73$ & $0.6971 \pm 0.0050$ & $3.43 \pm 0.12$ & \makecell{$0.02280$\\ $
					\pm 0.00015$} & $-0.972^{+0.11}_{-0.074}$ & $-0.37 ^{+0.66}_{-0.40}$ & $5.7^{+3.9}_{-2.8}$ & --& $ -868.66$ \\

				{\bf \(w_0 w_a\)CDM} & $69.77 \pm 0.72$& $0.6952 \pm 0.0049$& $3.38 \pm 0.13$& \makecell{$0.02274 \pm 0.00017$ }& $-0.868 \pm 0.052$& $-0.44^{0.23}_{-0.19}$& & & $-871.27$ \\

				{\bf $\Lambda$CDM} & $68.76 \pm 0.27$ & $0.6999 \pm 0.0034$ & & $0.02250\pm 0.00011$ &&&&& $-866.80$ \\
				
				\hline
				\multicolumn{10}{c}{\bf Data set: CC+PLA+DBAO+BBN+Union3}\\
				\hline
				\hline
				
				{\bf I, \((w_{\rm m} =-1.0)\)} & $72.0^{+1.4}_{-2.1}$ & $0.720^{+0.011}_{-0.014}$ &$3.25^{+0.16}_{-0.13}  $ & \makecell{$0.02266$\\$\pm 0.00017$} & $-3.6^{+1.4}_{-1.0}$ & $< 1.30   $ & $39.526^{-0.012}_{-13}$ & $16.6^{+2.5}_{-13}$ & $-51.44$ \\
				
				{\bf II, \((w_{\rm m} = -0.8)\)} &  $68.4^{+1.3}_{-0.49} $ & $0.6876^{+0.0087}_{-0.0041}$ & $3.337^{+0.099}_{-0.11}$ & \makecell{$0.02279$\\ $
					^{+0.00013}_{-0.00016}$} & $-0.36^{+0.33}_{-0.12}  $ & $27^{+8}_{-20}$ & $2.99^{+0.59}_{-2.0}$ & $> 35.2 $& $-58.28 $\\
				
				{\bf III, \((w_{\rm m} = -1.2)\)} & $68.0^{+1.5}_{-1.0} $ & $0.686^{+0.017}_{-0.013}$ & $3.18^{+0.12}_{-0.081}$ & \makecell{$0.02253$\\ $
					\pm 0.00017$} & $-1.47^{+0.46}_{-0.80}	$ & $-22^{+14}_{-22}  $ & $9.9^{+1.4}_{-2.3} $ & $31.9^{+3.3}_{-13}$& $-53.73$\\
				
				{\bf IV, \((w_{\rm m} = -1.0, b=0)\)} & $67.4\pm 1.2  $ & $0.6790\pm 0.0093$ & $3.24\pm 0.14$ & \makecell{$0.02261$\\ $
					\pm 0.00017$} & $-0.88^{+0.18}_{-0.13}$ & $-0.68^{+1.1}_{-0.86}$ & $5.3^{+2.8}_{-2.3} $ & --& $ -55.15$ \\

				{\bf \(w_0 w_a\)CDM} & $66.41 \pm 0.99$& $0.6793 \pm 0.0059$& $2.95 \pm 0.18$& \makecell{$0.02246 \pm 0.00020$ }& $-0.751 \pm 0.055$& $-0.79^{0.24}_{-0.21}$& & & $-54.22$ \\

				{\bf $\Lambda$CDM} & $68.59 \pm 0.29$ & $0.6979 \pm 0.0037$ & & $0.02247\pm 0.00011$ &&&&& $-51.89$ \\
				
				\hline
				\hline
				
			\end{tabular}
		}
		\caption{Marginalized constraints on the model parameters at the \(68\%\) confidence level for the different observational dataset combinations. Dashed entries indicate parameters that are not constrained by the corresponding dataset.  }
		\label{tab:best_fit_val}
		
	\end{table*}

	\begin{table*}
		\centering
		
		\resizebox{\textwidth}{!}{
			\begin{tabular}{l | ccccc cc c c r}
				\hline
				\multicolumn{11}{c}{\bf \boldmath Data set: CC+PLA+DBAO+BBN+$f\sigma_8$+$\sigma_8$+PP}\\
				\hline
				Model & \(H_0\) & $\Omega_{\rm de}$ & \(N_{\rm eff}\) & $\Omega_{\rm b}h^2$ & \(w_0\) & \(w_a\) & \(f\) & $b$ & $\sigma_8$ & $\log Z$\\
				\hline
				\hline
				{\bf   I, \((w_{\rm m} =-1.0)\)} & $66.21\pm 0.79$ & $0.6677^{+0.0055}_{-0.0062}$ & $3.26\pm 0.12$ & \makecell{$0.02265$\\ $
					\pm 0.00015$} & $1.59^{+0.41}_{-0.095}$ &  $< 2.55  $ &  $57^{+9}_{-7}  $ & $> 58.7  $ & $0.768\pm 0.020$  &$-884.95$\\

				{\bf    IV, \((w_{\rm m} = -1.0, b=0)\)} & $67.76 \pm 0.86$ & $0.6818\pm 0.0060$ & $3.24^{+0.16}_{-0.13}$ & \makecell{$0.02257$\\ $
					^{+0.00020}_{-0.00016}$} & $-0.859^{+0.11}_{-0.063}$ & $-0.28 ^{+0.89}_{-0.43}$ & $< 5.19$ & --& $0.766 \pm 0.020$ & $ -897.08$ \\

				{\bf  \(w_0 w_a\)CDM} & $67.47 \pm 0.88$& $0.6809 \pm 0.0059$& $3.18 ^{+0.15}_{-0.13}$& \makecell{$0.02256 \pm 0.00018 $ }& $-0.756 \pm 0.054$& $-0.77^{+0.24}_{-0.22}$& & & $0.766 \pm 0.020$& $-896.49$ \\
				
				{\bf   $\Lambda$CDM} & $68.49 \pm 0.29$ & $0.6967 \pm 0.0036 $& --& $0.02246 \pm 0.00011$&&&&& $0.766 \pm 0.020$& $-896.41$ \\
				
				\hline
				
				\multicolumn{11}{c}{\bf \boldmath Data set: CC+PLA+DBAO+BBN+$f\sigma_8$+$\sigma_8$+DES}\\
				\hline
				\hline
				{\bf   I, \((w_{\rm m} =-1.0)\)} &  $70.72^{+0.85}_{-1.1}$ & $0.7088^{+0.0050}_{-0.0075} $  & $3.28^{+0.13}_{-0.10} $  & \makecell{$0.02265$\\ $^{+0.00014}_{-0.00013}$} & $-3.5\pm 1.1$ & $< 1.32 $ & $67\pm 10 $ & $20\pm 11 $ & $0.770\pm 0.021 $ & $ -871.53$\\

				{\bf    IV, \((w_{\rm m} = -1.0, b=0)\)} & $68.50\pm 0.85$ & $0.6901\pm 0.0055$ & $3.22 \pm 0.14$ & \makecell{$0.02261$\\ $
					\pm 0.00018$} & $-0.932^{+0.11}_{-0.075}$ & $-0.12 ^{+0.69}_{-0.36}$ & $< 6.18   $ & --& $0.769 \pm 0.020$ &$ -876.89 $ \\

				{\bf   \(w_0 w_a\)CDM} & $68.25 \pm 0.87$& $0.6878 \pm 0.0056$& $3.22 \pm 0.14$& \makecell{$0.02257 \pm 0.00019 $ }& $-0.825 \pm 0.056$& $-0.58^{0.25}_{-0.21}$& & & $0.767 \pm 0.020$ & $-877.06$\\

				{\bf   $\Lambda$CDM} & $68.56 \pm 0.28$  & $0.6975 \pm 0.0036$ & & $0.02246 \pm 0.00011$&&&&& $0.767 \pm 0.020$& $-871.43$\\

				\hline
				\hline

		\end{tabular} }
		
		\caption{Marginalized constraints on the model parameters at the \(68\%\) confidence level for the different observational dataset combinations (includes $f\sigma_8$). Dashed entries indicate parameters that are not constrained by the corresponding dataset. }
		\label{tab:best_fit_val2}
	\end{table*}
	
	\begin{figure*}
		\centering
		\includegraphics[scale=0.43]{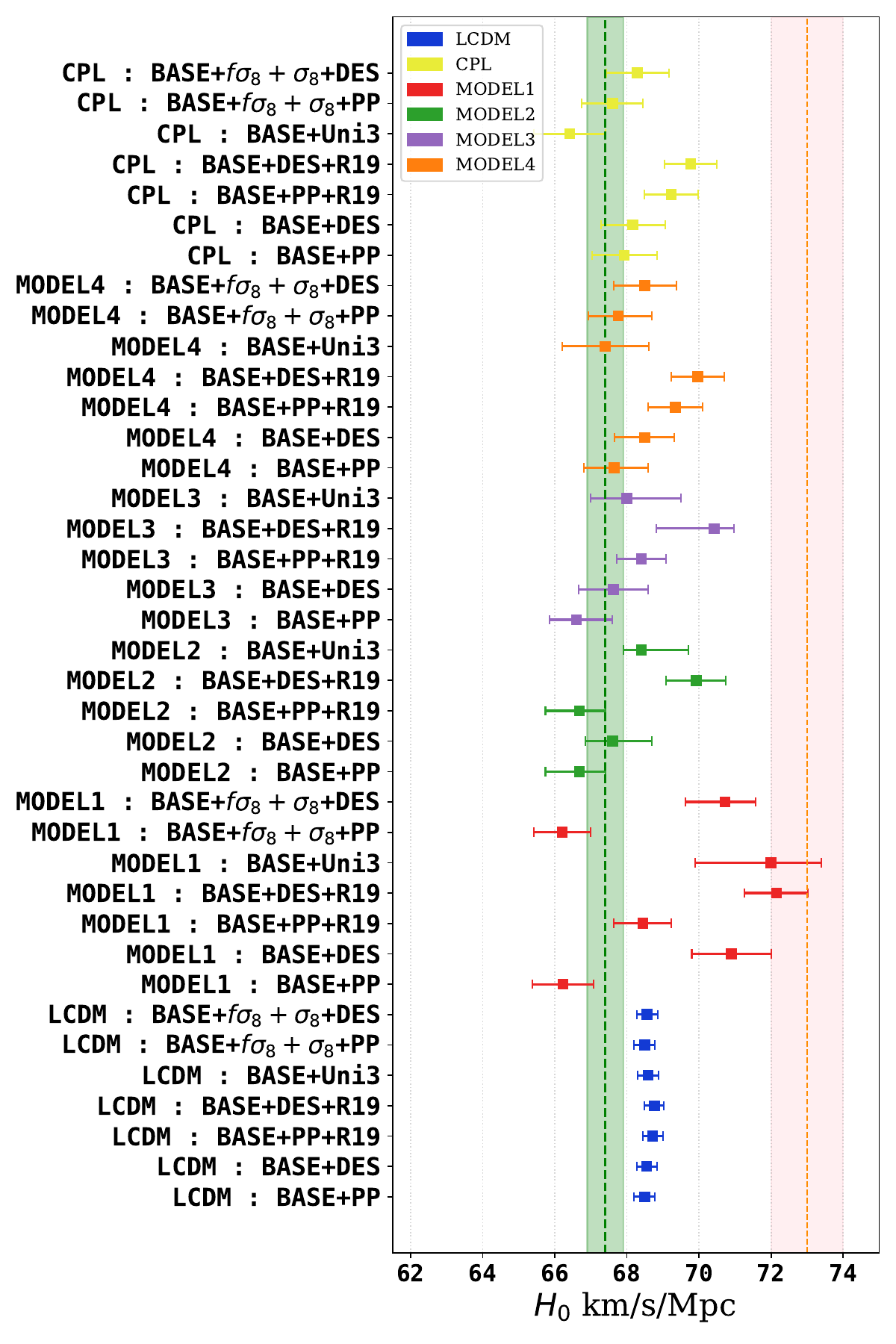}
		\includegraphics[scale=0.43]{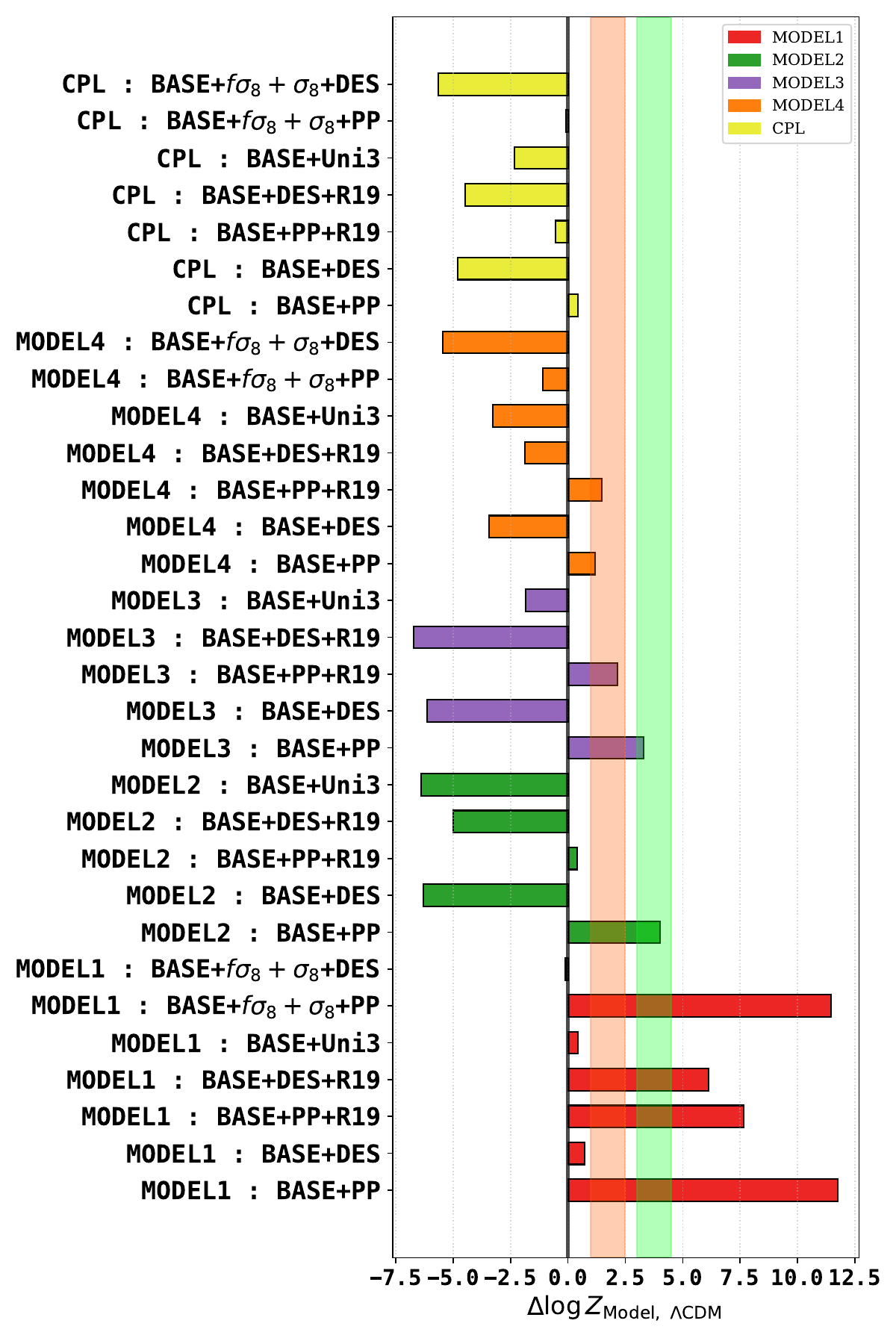}
		\caption{The values of $H_0$ (left) at 68\% confidence level and the difference in log evidence,
			$\Delta \ln Z \equiv \ln Z_{\rm Model}-\ln Z_{\Lambda{\rm CDM}}$ (right).}
		\label{fig:H0_best}
	\end{figure*}

	%%%%%%%%%%%%%%%%%%%%%%%%%%%%%%%%%%%%%%%%%%%%%%%%%%

	% Don't change these lines
	% \bsp	% typesetting comment
	\label{lastpage}
\end{document}